\documentclass[aps,prb,reprint,superscriptaddress]{revtex4-2}
\usepackage{graphicx}
\usepackage{amsmath}
\usepackage{amssymb}
\usepackage{natbib}
\usepackage{xcolor}
\usepackage{hyperref}
\usepackage{upgreek}  % 导言区添加
\usepackage{color}
\begin{document}
\title{Out-of-plane ferroelectricity, magnetoelectric coupling and persistent spin texture in two-dimensional multiferroics} 
\author{Ying Zhou}
\email{These authors contributed equally to this work}
\author{Cheng-Ao Ji}
\email{These authors contributed equally to this work}
\affiliation{Key Laboratory of Quantum Materials and Devices of Ministry of Education, School of Physics, Southeast University, Nanjing 211189, China}
\author{Shuai Dong}
%\email{sdong@seu.edu.cn}
%\affiliation{Key Laboratory of Quantum Materials and Devices of Ministry of Education, School of Physics, Southeast University, Nanjing 211189, China}
\author{Xuezeng Lu}
\email{xuezenglu@seu.edu.cn}
\affiliation{Key Laboratory of Quantum Materials and Devices of Ministry of Education, School of Physics, Southeast University, Nanjing 211189, China}

\begin{abstract}
Two-dimensional multiferroics with out-of-plane ferroelectricity hold significant promise for miniaturized magnetoelectric spin-orbit transistors, yet systems combining robust ferroelectricity and strong magnetoelectric coupling are exceedingly rare. Here, we demonstrate that epitaxial strain stabilizes out-of-plane ferroelectricity in exfoliated two-dimensional Ruddlesden-Popper derivatives. The hybrid improper ferroelectric $Pc$ phase transitions to a competing $P2_1$ phase with purely in-plane polarization upon switching, accompanied by a 90$^\circ$ rotation of weak ferromagnetism. Crucially, the $Pc$ phase exhibits altermagnetism, while $P2_1$ displays full-Brillouin-zone band splitting, with persistent spin textures rotating 90$^\circ$ at the phase boundary. This work establishes a pathway to engineer two-dimensional multiferroics that integrate vertical polarization, magnetoelectric coupling, and switchable spin textures—key features for next-generation spintronic devices.
\end{abstract}

\maketitle

\section{Introduction}
Ferroelectric materials have attracted considerable attention for modern technologies due to their bistable polarization states, making them promising candidates for nonvolatile memory applications \cite{Chang2016,Dong2019,Banerjee2022}.The demand for miniaturized devices requires stabilization of out-of-plane ferroelectricity (i.e., oriented along the thin-film growth direction), thus, their associated functionalities such as multiferroicity, magnetoelectricity and ferroelectric switchable spin texture can persist as the film thickness decreases \cite{Dawber2005,Scott2007}. Nevertheless, conventional ferroelectrics possess an intrinsic critical thickness below which the out-of-plane polarization disappears, fundamentally restricting their implementation in ultrathin devices \cite{Junquera2003,Spaldin2004}.

Recently, the study of two-dimensional (2D) materials has found promising mechanisms for exploring out-of-plane ferroelectricity in the ultrathin film and some of them have been realized in experiments \cite{Song2022,Yasuda2021,PhysRevLett.132.086802}. For instance, 2D CuInP$_2$S$_6$ was experimentally confirmed to host vertical ferroelectricity, driven by the displacement of Cu and In ions along the crystallographic $c$-axis, representing the first confirmed 2D system with intrinsic out-of-plane polarization \cite{Belianinov2015,Liu2016}. First principles calculations subsequently revealed that interlayer sliding in the materials with van der Waals (vdW) bilayers (e.g., BN, AlN, ZnO, MoS$_2$, GaSe, etc.) can lead to vertical polarization \cite{Li2017}. Experimentally, spontaneous out-of-plane polarization has been observed in multi-layer WTe$_2$ ($2–3$ layers) \cite{Fei2018}. Additionally, 2D $\alpha$-In$_2$Se$_3$ has been predicted to possess coupled in-plane and out-of-plane polarizations arising from Se atomic displacements \cite{Ding2017,Zhou2017}.

So far, the exploration of 2D materials has largely centered on vdW systems, with very few examples exhibiting both robust out-of-plane ferroelectricity and magnetoelectric (ME) coupling. In contrast, 3D multiferroics with strong ME coupling typically rely on interconnected transition-metal coordination geometries—such as edge-, corner-, or face-sharing octahedra, tetrahedra, or square pyramids  \cite{Hu2019,Spaldin2019,Spaldin2017,Dong2015}. This structural connectivity, while enabling strong ME effects in 3D systems, poses a significant challenge for reducing these materials to their 2D limit, as ultrathin films often lose their out-of-plane ferroelectric properties. However,  the experimental breakthroughs in synthesizing 2D monolayers from both layered and nonlayered transition-metal oxides have opened new avenues for investigating ferroelectricity in atomically thin systems \cite{Nicolosi2013,Coleman2011,Balan2018,Ji2019}. Capitalizing on these advances, there is growing interest in leveraging the dense transition-metal networks found in 3D perovskite oxides to engineer 2D materials that retain both out-of-plane ferroelectricity and ME coupling—a critical step toward realizing functional ultrathin multiferroics.

In this work, by using first-principles calculations and group theory analysis, we demonstrate that the structure containing two perovskite layers derived from the quasi-2D Ruddlesden-Popper (RP) Ca$_3$Mn$_2$O$_7$ can have out-of-plane ferroelectricity under the expitaxial strain $x \neq y$ ($x$, $y$: the in-plane lattice parameters). The stabilized polar phase has the $Pc$ symmetry and the ferroelectricty emerges from the cooperative octahedral rotations and tilts, thus,  is the hybrid improper ferroelectricity. Additionally, we predict a phase transition from the $Pc$ to $P2_1$ under the epitaxial strain, where the $P2_1$ phase has the close energy to the $Pc$ phase. Meanwhile, there is a change on the  direction of the weak ferromagnetism (wFM) at the phase transition. More interestingly, the $Pc$ phase has the altermagnetic ordering and $P2_1$ phase has a full-Brillouin-zone (FBZ) band splitting. Both phases have the persistent spin textures (PST)  that can be transformed to each other through changing the out-of-plane hybrid improper ferroelectricity (OP-HIF). In our studied PST, the quadratic terms of the momentum vectors dominate the spin-orbit field, contrasting to the previously reported PST with the linear or cubic terms of the momentum vectors, and a 90$^\circ$ rotation of the PST can occur at phase transition. Moreover, we also find the $Pc$ phase in the RP halides, where the epitaxial strain stabilizes the $Pc$ phase. These results establish the design principles for discovering materials with out-of-plane ferroelectricity and strong ME coupling based on the perovskite structures.

\section{Computational Methods}
The first-principles calculations based on density functional theory (DFT) were performed using the projector-augmented wave method, as implemented in the Vienna \emph{ab initio} simulation package \cite{Kresse1996,Kresse19961,BLOCHL1994}. The Perdew-Burke-Ernzerhof functional was used as the exchange-correlation functional \cite{Kresse1999,WANG1991}. A vacuum space of $25$ {\AA} was added to avoid interaction between neighboring periodic images. We used a cutoff energy of $600$ eV for the plane-wave bases, and a $\varGamma$-centered $5\times5\times1$ $k$-point mesh for the Brillouin zone integration. The in-plane lattice constants and internal atomic coordinates of each structural phase were relaxed until the Hellman-Feynman force on each atom is less than $0.005$ eV/{\AA}. A convergence threshold of $10^{-6}$ eV was used for the electronic self-consistency loop. To describe correlated $3d$ electrons of Mn, the GGA+$U$ method is applied \cite{Dudarev1998}, and the $U_{eff}$ are $3.5$ eV and $6$ eV for 2D Ca$_3$Mn$_2$O$_7$ and 2D KAg$_2$Mn$_2$Cl$_7$, respectively. Phonon band structures were calculated using density functional perturbation theory (DFPT). The phonon frequencies and corresponding eigen-modes were calculated on the basis of the extracted force-constant matrices with a $4\times4\times1$ supercell, as implemented in the PHONOPY code \cite{Togo2015}. The polarization was calculated by using the Berry phase method \cite{KINGSMITH1993}. The energy barrier of different ferroelectric switching paths was determined by using the climbing nudged elastic band method \cite{Sheppard2012}.The ISOTROPY tool \cite{Campbell2006} was used to help with the group-theoretical analysis.

\section{Results and Discussion}
In bulk Ca$_3$Mn$_2$O$_7$, first-principles calculations predicted the existence of hybrid improper ferroelectricity, in which the polarization is driven by a trilinear coupling interaction (~$PQ_RQ_T$) among the polarization ($P$), in-phase octahedral rotation ($Q_R$) and out-of-phase tilt ($Q_T$) \cite{PhysRevLett.106.107204}. Without this anharmonic mode-mode interaction, polarization cannot be condensed by itself. Through the trilinear coupling interaction, a ME coupling mechanism was elucidated within the G-type antiferromagnetic (AFM) structure, where the wFM is induced by the Dzyaloshinskii-Moriya (DM) interaction that is proportional to $Q_T$, thus, a simultaneous change of polarization and $Q_T$ can lead to a switchable wFM. But if another polarization switching through a simultaneous change of $P$ and $Q_R$ occurs, there is no control of the wFM. In fact, these two polarization switching paths usually compete with each other, which leads to the ME coupling remaining unrealized in experiments so far. This is a dilemma that most type-I multiferroics encounter. For example, BiFeO$_3$ also has two polarization switching paths \cite{ChuYing-Hao2008, ZhaoT2006}. In one path, there is only change of polarization and out-of-phase rotations around the pseudocubic [111] direction cannot be changed. In another path, both polarization and out-of-phase rotations can be switched. Only the latter one can have simultaneous change of wFM and polarization \cite{HeronJT2014, Shutong2020, PhysRevB.94.104105, PhysRevB.107.054108, PhysRevB.105.075408, PhysRevB.105.054434}. 

\begin{figure}
	\centering
	\includegraphics*[width=0.49\textwidth]{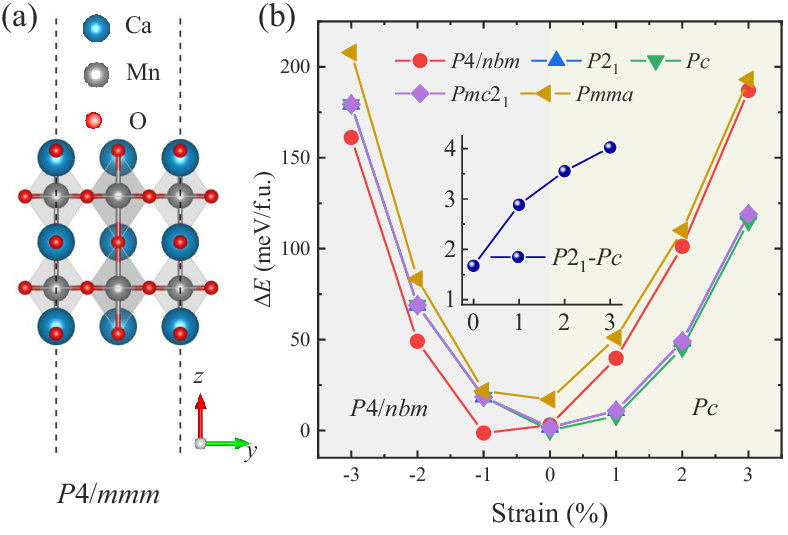}
	\caption{(a) The high-symmetry $P4/mmm$ structure of the 2D-RP structure of Ca$_3$Mn$_2$O$_7$. (b) The energies of several 2D-RP structures of Ca$_3$Mn$_2$O$_7$ as a function of the epitaxial strain. The energies are calculated with respect to that of $P4/nbm$ at -1 \% strain. The insert shows the energy differences of the $P2_1$ phase relative to the $Pc$ phase  as a function of the tensile strain ranging from 0\% - 3\%.}
	\label{Fig1} 
\end{figure}

Although many studies reported the layered oxides with the polar ground state phase such as in the RP phase, Aurivillius phase, Dion-Jacobson phase, the realization of ME couplings in these materials in experiments are still difficult as mentioned above. To address the above dilemma, the investigations on the RP halides provided insight into the solutions: (1) searching for the polar ground state phase with low symmetry. For example, the $Pc$ phase can be stabilized in KAg$_2$Mn$_2$Cl$_7$ with G-type AFM ordering \cite{Lu2023}, where both $Q_R$ and $Q_T$ can control the wFM because that the only one mirror symmetry cannot lead to a perfect cancellation of the weak antiferromagnetism controlled by $Q_R$. Moreover, the $Pc$ phase of KAg$_2$Mn$_2$Cl$_7$ has the OP-HIF, which is the secondary mode and induced by two out-of-phase ($Q_{R1}$) and in-phase ($Q_{R2}$) octiahedral rotations, i.e., $PQ_{R1}Q_{R2}$. Then, the switching of the OP-HIF can lead to a change of either $Q_{R1}$ or $Q_{R2}$. No matter which one is changed, the simultaneous change of the OP-HIF and wFM can occur. This is not related to the polarizaiton switching processes. (2) Finding reliable polar-to-nonpolar phase transition with the simultaneous change of the polarization and magnetism, which was found in (Rb,K)$_3$Mn$_2$Cl$_7$ with G-type AFM ordering and there are two kinds of $Q_T$ in the two phases that result in the different directions of wFM. Unfortunately, the polar-to-nonpolar phase transition with the simultaneous change of the polarization and magnetism in (Rb,K)$_3$Mn$_2$Cl$_7$ was only realized by varying the temperature, that is, thermally controlled ME coupling \cite{Zhu2024}.

It is interesting to continue searching the ME multiferroics, especially, one has out-of-plane ferroelectricity, below $1$ nanometer in the RP materials. Previous study showed that the double-perovskite-layer structure derived from Ca$_3$Mn$_2$O$_7$ can have a $Pmc2_1$ polar ground state structure with the in-plane polarization and G-type AFM orderings under biaxial strain \cite{ZY2021}. Although the ME coupling was also predicted in the 2D structure, that still relied on the trilinear coupling interaction, thus, may be invalid when the polarization switching through the change of $Q_R$ occurs. Our DFT calculations indicate that the $Pc$ phase found in the RP halides can be stabilized in the double-perovskite-layer structure (hereafter, it is called 2D-RP structure) derived from Ca$_3$Mn$_2$O$_7$ by the epitaxial strain with $x=1.004y$ (see Fig. \ref{Fig1} and Fig. S1 in the Supplemental Material \cite{SM}). The form of the epitaxial strain in our study may be a reasonable description of a strained film for some materials where a buffer layer usually exists between film and substrate, resulting in imperfect registry \cite{Hatt2009}. Under tensile strain, the energy difference between the metastable $P2_1$ and $Pc$ phases increases from approximately 1.5 to 4.5 meV/f.u. as the strain rises from 0\% to 3\%, as shown in Fig. \ref{Fig1} (b). And the previously found $Pmc2_1$ phase becomes $P2_1$ phase with the epitaxial tensile strain. The energy difference between $Pc$ and $P2_1$ phases can become large with increasing the $x/y$ ratio (e.g., $11$ meV/f.u. at $1\%$ strain and $x=1.04y$). Therefore, the $x/y$ ratio may be an effective factor to engineer the out-of-plane ferroelectricity in the 2D-RP structure. We also find the $Pc$ ground state phase in the 2D-RP structure of the RP halides (see Fig. S2, Fig. S3 and Table S3 \cite{SM}).

\begin{figure}
	\centering
	\includegraphics*[width=0.45\textwidth]{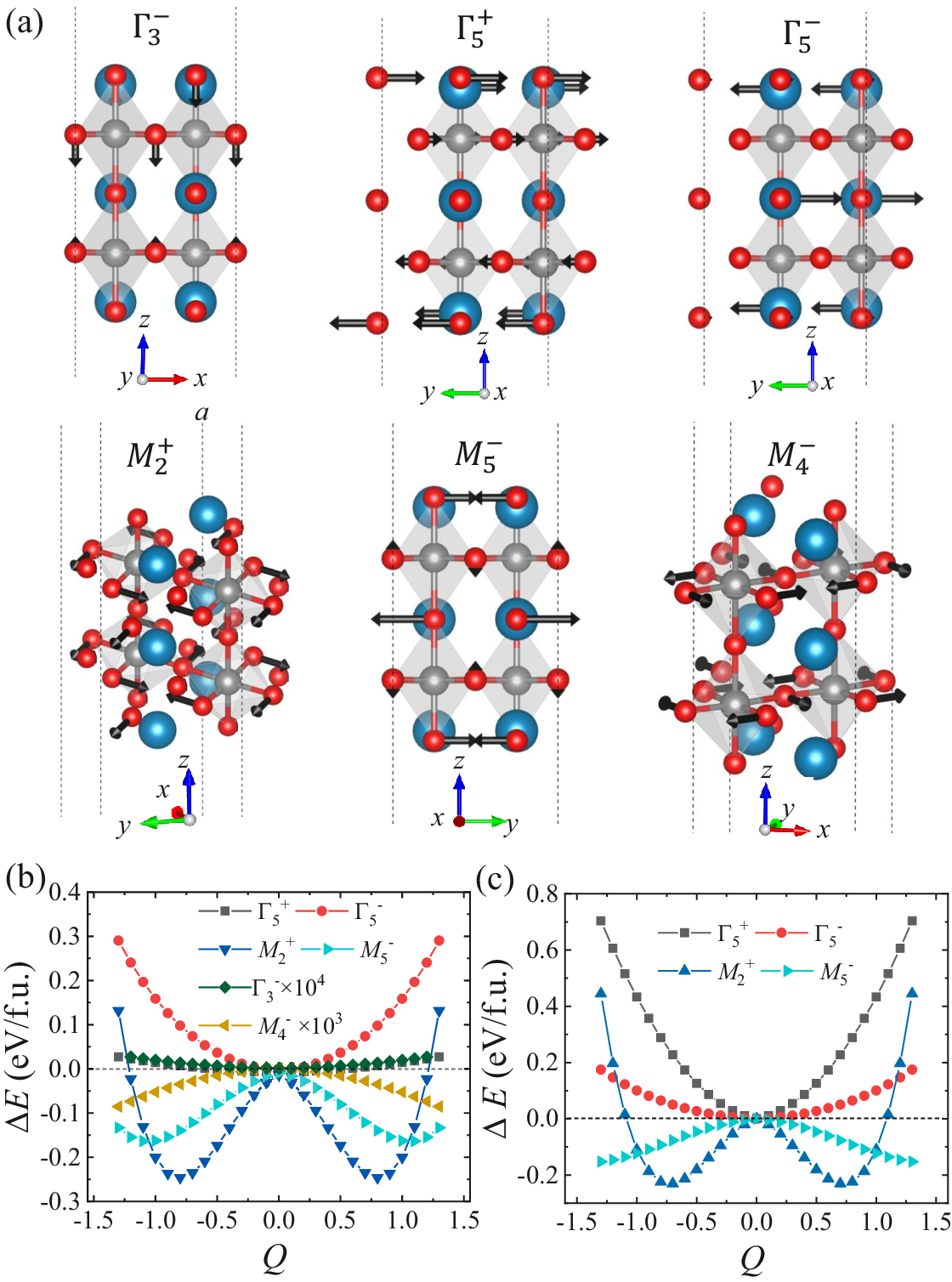}
	\caption{(a) The main atomic displacement patterns in the $Pc$ phase of the 2D-RP structure of Ca$_3$Mn$_2$O$_7$, represented by the designated irreps of the $P4/mmm$ phase. (b) Shows the energy of each atomic displacement shown in the high-symmetry $P4/mmm$ phase where the mode amplitude $Q$ is scaled by the original amplitude of the corresponding atomic displacement in the $Pc$ structure. The energy is computed relative to the high-symmetry $P4/mmm$ phase. Energy gains for the $M_4^-$ and $\Gamma_3^-$ modes are magnified by the factors of $10^3$ and $10^4$, respectively.} (c) The energy of each atomic displacement in the $P2_1$ phase versus the mode amplitude $Q$.
	\label{Fig2}
\end{figure}

Next, the mode decomposition was performed for $Pc$ phase with respect to high-symmetry $P4/mmm$ phase [see Fig. \ref{Fig1} (a)]. As shown in Fig. \ref{Fig2}(a), there are six main phonon modes, which are an in-plane polar mode having irreducible representation (irrep) $\Gamma_5^-$, an out-of-plane polar mode $\Gamma_3^-$, an antiferroelectric (AFE) mode $\Gamma_5^+$, an in-phase octahedral rotational (OOR$_i$) mode $M_2^+$ around $z$ direction, an out-of-phase octahedral rotational (OOR$_o$) mode $M_4^-$ around $z$ direction and an out-of-phase octahedral tilt (OOT) mode $M_5^-$ around $x$ direction. Then, two trilinear coupling interactions can be obtained, which are $Q_{\Gamma_5^-} Q_{M_2^+} Q_{M_5^-}$ and $Q_{\Gamma_3^-} Q_{M_2^+} Q_{M_4^-}$. Since only the OOR$_i$ and OOT have large energy gain [see Fig. \ref{Fig2}(b)], the in-plane and out-of-plane polarizations are induced by the trilinear coupling interactions $Q_{\Gamma_5^-} Q_{M_2^+} Q_{M_5^-}$ and $Q_{\Gamma_3^-} Q_{M_2^+} Q_{M_4^-}$, respectively. Because the OP-HIF has the ability to overcome the depolarized field \cite{Lu2023,PhysRevLett.102.107601}, thus the $Pc$ phase can be stable in the 2D-RP structure. But the depolarized field still have effects on the OP-HIF, where the remanent polarization with the existence of the depolarized field will be small in the ultrathin film \cite{PhysRevLett.112.127601}. The computed polarization of the $Pc$ phase in the 2D-RP structure is ($11.03$, $0$, $0.14$) $\mu$C/cm$^2$, in which the in-plane hybrid improper ferroelectricity (IP-HIF) is comparable to the other RP materials in bulk \cite{Akamatsu2014,PhysRevB.94.104105}. In the $P2_1$ phase, there are four main phonon modes, which are $\Gamma_5^-$,  $\Gamma_5^+$,  $M_2^+$ and $M_5^-$ [see Fig. \ref{Fig2}(c) and Table S2 \cite{SM}]. The $M_5^-(0,a)$ has different order parameter directions (OPDs) from $M_5^-(a,0)$ of $Pc$ phase and the $P2_1$ phase only has IP-HIF $(0, 10.28, 0)$ $\mu$C/cm$^2$ that arises from $Q_{\Gamma_5^-} Q_{M_2^+} Q_{M_5^-}$. 

\begin{figure}
	\centering
	\includegraphics*[width=0.45\textwidth]{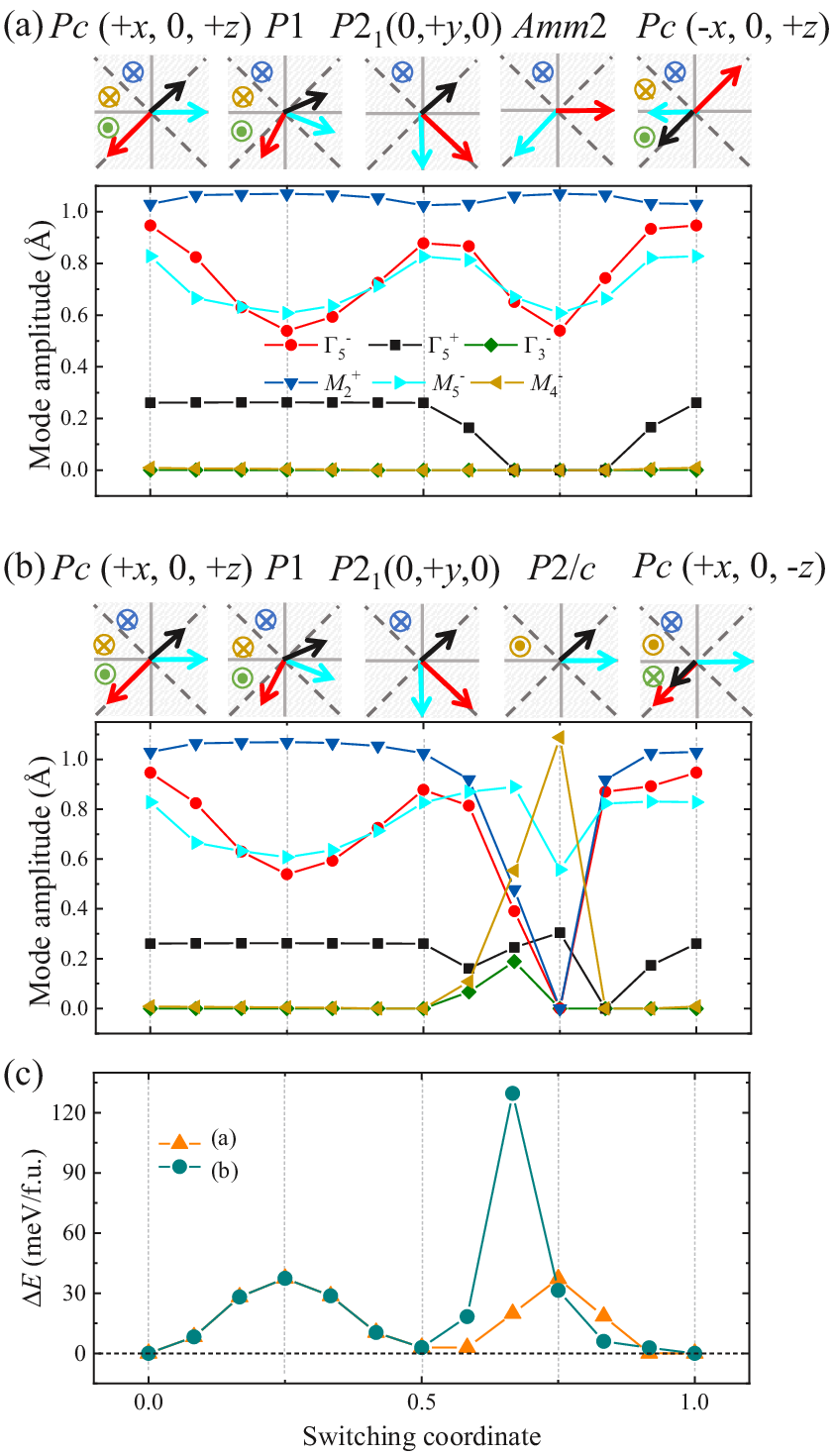}
	\caption{Ferroelectric switching paths with the switched (a) the IP-HIF and the OP-HIF (b) of the $Pc$ for 2D-RP structure of Ca$_3$Mn$_2$O$_7$ at 1$\%$ tensile strain, where $x$, $y$, and $z$ denote the polarization components along the three Cartesian axes. Symbols in upper panels indicate the evolution of order parameters during polarization switching: arrows for 2D order parameters ($\Gamma_5^+$ [black], $\Gamma_5^-$ [red], $M_5^-$ [cyan]); $\otimes$/ $\odot$ for 1D order parameters ($\Gamma_3^-$ [green], $M_2^+$ [blue], $M_4^-$ [gold]). The bottom part shows the absolute amplitudes as a function of the switching coordinate, obtained from nudged elastic band calculations. (c) The computed energy barriers for the paths shown in (a)–(b).}
	\label{Fig3} 
\end{figure}

In our nudged elastic band (NEB) calculations \cite{Sheppard2012} [see Fig. \ref{Fig3} (c)], we find that the IP-HIF can be switchable with a barrier of $40$ meV/f.u., while the switching barrier is ~$120$ meV/f.u. for OP-HIF indicating that it may be hard to be switchable. In our two-step polarization switching paths, the $Pc$ phase may be transformed to the $P2_1$ phase by tuning the OP-HIF by applying out-of-plane electric field, where our computed barrier of $40$ meV/f.u. and OP-HIF of $0.14$ $\mu$C/cm$^2$ are comparable to those in the vdW materials \cite{Guan2020, Ma2021}. The 90° twin domains of the $Pc$ and $P2_1$ phases are also considered in the polarization switching, as it usually leads to a low energy barrier \cite{PhysRevB.94.104105, Zhu2024}. We find that the $Pc$ with IP-HIF along $y$ direction and OP-HIF (twin domain of $Pc$) and $P2_1$ with IP-HIF along $x$ direction (twin domain of $P2_1$) cannot exist with $x=1.004y$, since the twin domain of $Pc$ and twin domain of $P2_1$ are unstable and will be relaxed to $P2_1$ and $Pc$, respectively. Thus, the slightly anisotropic $x$ and $y$ under epitaxial strain will help the transformation between the $P2_1$ with IP-HIF along the $y$ direction and $Pc$ with the IP-HIF along the $x$ direction by tuning the OP-HIF. Besides, the $Pc$-to-$P2_1$ transition can also be accessible through low in-plane electric field as the IP-HIF is $11.03$ $\mu$C/cm$^2$.

Then, we consider the modes changes in the ferroelectric switching paths \cite{PhysRevB.94.104105, PhysRevB.107.054108, PhysRevB.105.075408, PhysRevB.105.054434}. In the first path shown in Fig.~\ref{Fig3}(a), the initial $Pc$ phase exhibits the polarization along $(+x, 0, +z)$. This state evolves through an intermediate $P2_1$ phase with the	polarization along $(0, +y, 0)$, finally reaching the $Pc$ phase with the polarization along $(-x, 0, +z)$ station. In this path, the $M_2^+$ rotational mode is kept and the $\Gamma_5^-$ polar mode and $M_5^-$ tilt mode are switched, which is consistent with the existence of the trilinear coupling $\sim$ $Q_{\Gamma_5^-} Q_{M_2^+} Q_{M_5^-}$. In the second ferroelectric reversal path shown in Fig.~\ref{Fig3}(b), the $Pc$ phase has the switched polarization polarized along $(+x, 0, -z)$. Along this switching path, the $\Gamma_3^-$  polar mode and $M_4^-$ mode are changed and the $M_2^+$ is still kept ,which again fulfills the trilinear coupling interaction of $Q_{\Gamma_3^-} Q_{M_2^+} Q_{M_4^-}$. In both switching paths, the $M_2^+$ rotational mode is unchanged because of its large energy gain in stabilizing the ground state $Pc$ phase, and this can facilitate the switching processing with the low energy barrier.

\begin{figure}
	\centering
	\includegraphics*[width=0.45\textwidth]{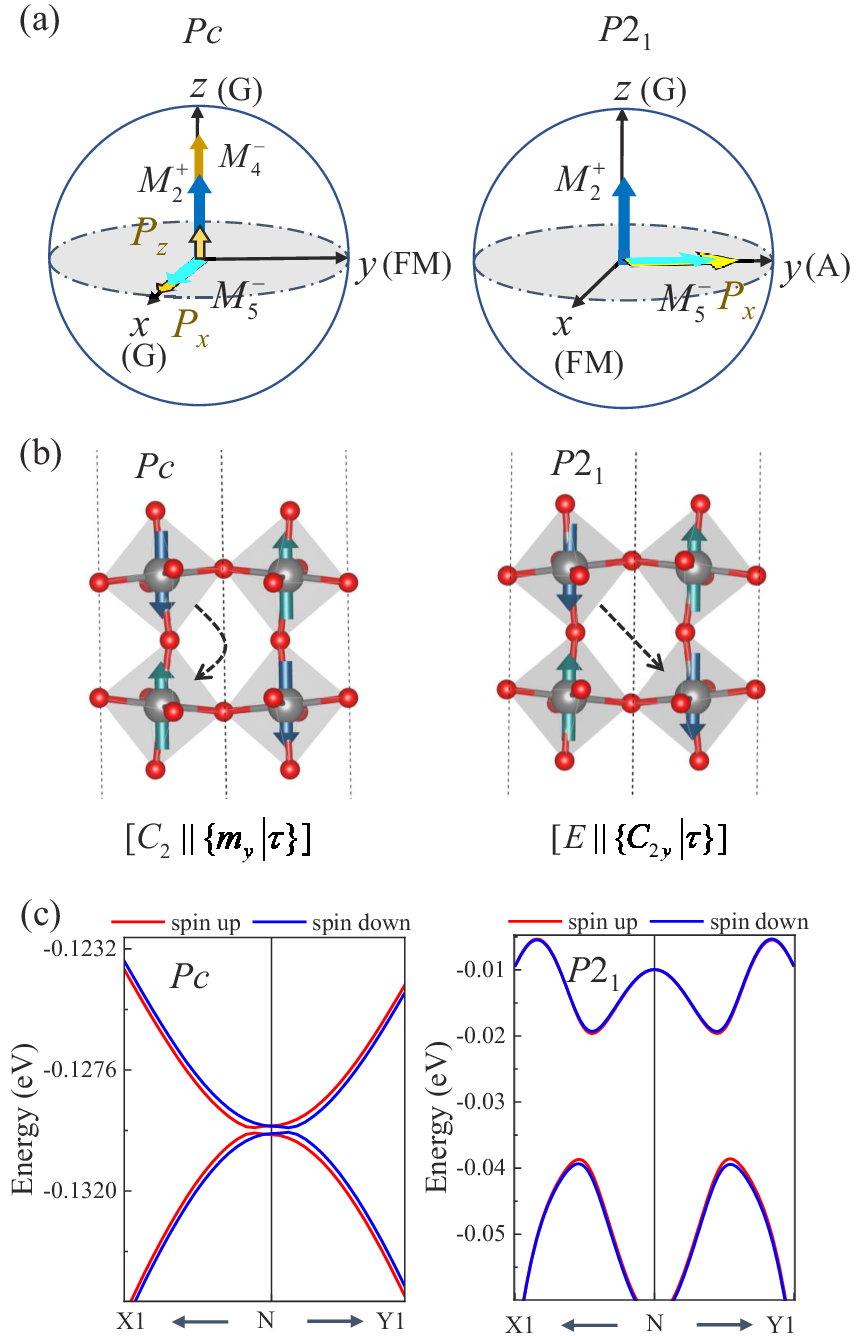}
	\caption{(a) Magnetic symmetry analysis for the initial $Pc$ ($+x, 0, +z$) and the intermediate $P2_1$ ($0, +y, 0$) phases. In the $Pc$ phase, the magnetic configurations along the three orthogonal axes are: G-type AFM along the $x$-axis, FM along the $y$-axis, and G-type AFM along the $z$-axis. In the $P2_1$ phase, the configurations correspond to (FM, A-type AFM, G-type AFM) along the three axes. Blue/golden arrows denote the in-phase ($M_2^+$)/out-of-phase ($M_4^-$) rotation mode, cyan arrows represent the tilt mode ($M_5^-$), and yellow arrows indicate the direction of the polarization. (b) Spin ordering with the antiparallel/parallel spins mapped by the spin group operations for $Pc$ and $P2_1$ phases. (c) The band structures without SOC of the $Pc$ and $P2_1$ phases at 1\% strain, the VBM is set to 0 eV. The kpoints are N (0.5, 0, 0), X1 (0.2, 0.2, 0), Y1 (0.2, -0.2, 0).}
	\label{Fig4} 
\end{figure}

Next, we focus on the magnetic properties. The magnetic structure of Mn$_4^+$ ions has G-type magnetic ordering for the main spin direction, which has been demonstrated in both the 2D-RP with the biaxial strain ($x=y$) and bulk Ca$_3$Mn$_2$O$_7$ structures \cite{ZY2021, ChenBuHang2021, Benedek2011, PhysRevLett.114.035701, LiuMeifeng2018}. Our DFT+$U$+SOC (spin-orbital coupling) calculations show that the magnetocrystalline anisotropy energy (MAE) is along the out-of-plane direction for the $Pc$ and $P2_1$ phases, as shown in Fig. S4 \cite{SM}. The wFM are computed to $(0.001, 0.038, 0.000) \mu_B$/u.c. and $(0.042,0.002, 0.000) \mu_B$/u.c. for the $Pc$ and $P2_1$ phases, respectively, where each unit cell (u.c.) contains four magnetic atoms. The computed magnetic configurations by including SOC are consistent with the symmetry-adapted configurations of (G, FM, G) for the $Pc$ phase (the magnetic point groups (MPG) is $m$) and (FM, A, G) for the $P2_1$ phase (the MPG is $2'$), as shown in Fig. \ref{Fig4}(a). 

By symmetry analysis \cite{PhysRevX.12.040501}, we discovered that the spin up and spin down sublattices are connected by the spin space symmetry [$C_2$$\|$\{$m_y$$\vert$$\uptau$\}] and there are no inversion and translational symmetries between the antiparallel spins in $Pc$ phase [see Fig. \ref{Fig4} (b)]. Hence, we can confirm that the G-type AFM of $Pc$ phase is altermagnetic, consistent with our band structure calculations without SOC [see Fig. \ref{Fig4} (c)].
	
The band splitting of $Pc$ phase without SOC between the spin-up and spin-down bands near the N point shown in Fig. \ref{Fig4} (c) is about 0.18 meV. By considering the SOC, the band splitting becomes 0.20 meV. Although the band splitting is small, the variation of the band splitting with the momentum vector $k$ ($\sim$$\Delta$$E$/$\Delta$$k$, $\Delta$$E$=$E$$_{\textit{spin­-up}}$-$E$$_{\textit{spin­-down}}$) with and without SOC are about 22.16 meV·\AA {} and 20.68 meV·\AA {} in the $Pc$ phase. This band splitting strength $\sim$$\Delta$$E$/$\Delta$$k$ is comparable to those in the quantum-wells or LaAlO$_3$/SrTiO$_3$ heterointerfaces that typically exhibit $1-10$ meV·\AA \cite{PhysRevB.96.161303, PhysRevLett.104.126803,  Choe2019}. These systems have been extensively studied for their spin textures in experiments.
%PhysRevLett.90.032301, 	
Furthermore, recent experimental advances have already demonstrated the observation of the altermagnetic phases in the materials such as $\alpha$-MnTe and RuO$_2$, using techniques like angle-resolved photoemission spectroscopy (ARPES) and magnetic circular dichroism \cite{Fedchenko2024, PhysRevLett.132.036702}. These results confirm that altermagnetism can be achieved and detected in real systems, indicating strong potential for experimental realization in other candidate materials as well.

\begin{figure}
	\centering
	\includegraphics*[width=0.5\textwidth]{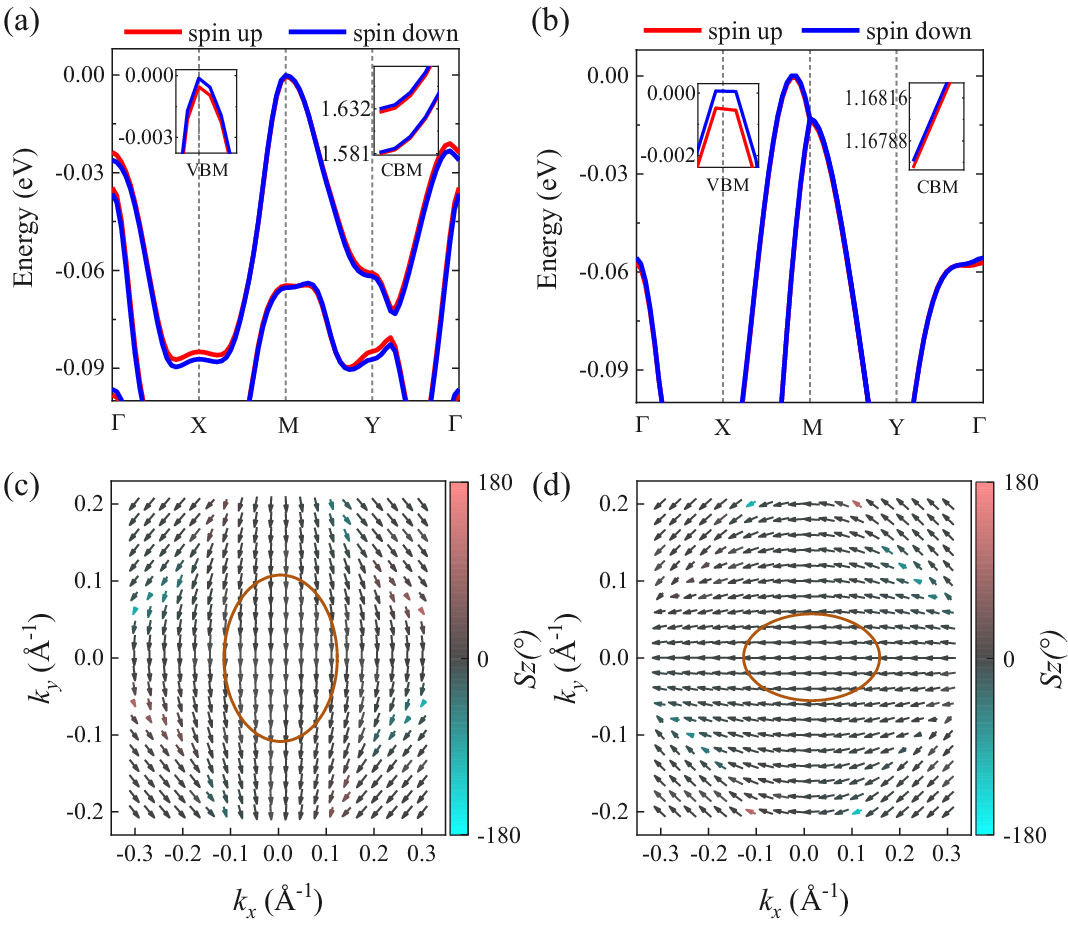}
	\caption{(a) and (b) show the band structures without SOC of the $Pc$ and $P2_1$ phases at 1\% strain with an applied electric field of 0.25 V/$\mathrm{\AA}$, the VBM is set to 0 eV. (c) and (d) The PST in the two phases at 1\% strain of the 2D-RP structure of Ca$_3$Mn$_2$O$_7$ with SOC at $E$= 0 V/$\mathrm{\AA}$. The red circle labels the maximum regions of PST (where the spin deviation is less than $5^\circ$) when the energy gradient is considered.}
	\label{Fig5} 
\end{figure}

For the $P2_1$ phase, there is no any symmetry between the antiparallel spins and the parallel spins are connected by the symmetry [$E\|$\{$C_{2y}$$\vert$$\uptau$\}] [see Fig. \ref{Fig4} (b)], therefore, the $P2_1$ should have FBZ band splitting without SOC. This is demonstrated in our band structure calculations without SOC [see Fig. \ref{Fig4} (c)]. To make the FBZ band splitting large in the $P2_1$ phase, an electric field ($E$) of 0.25 V/$\mathrm{\AA}$ is applied to be along the $2$-fold rotation in which the $P2_1$ symmetry is kept, following a method based on a Landau model originally used to compute the linear ME response \cite{PhysRevLett.103.267205}. Then, a more visible FBZ band splitting can be obtained [see Fig. \ref{Fig5}(b)]. If the electric field is applied to be perpendicular to the mirror plane in the $Pc$ phase, there will be no symmetry among antiparallel spins and the FBZ band splitting without SOC can also be observed [see Fig. \ref{Fig5}(a)]. Besides, the FBZ band splitting without SOC can be tuned by the OP-HIF through the $Pc$-to-$P2_1$ phase transition, leading to a change on the amplitude of FBZ band splitting around conduction band maximum (CBM) that can be related to the carrier/spin transport properties \cite{Madsen2018, PhysRevB.104.075433}. In the $Pc$ phase, the effective mass around the CBM is $1.8m_0$ (electron mass) along the $\Gamma$-Y direction, while in the $P2_1$ phase, it reduces to $0.7m_0$ along $\Gamma$-X (see Fig. S8 \cite{SM}). Our findings take advantages over the previous studies by asymmetric ion substitution in bulk or in the other 2D materials where the full-Brillouin-zone band splitting cannot be tunable \cite{PhysRevLett.133.216701, PhysRevLett.134.116703}.  

Previous theory combined with experiment has demonstrated the existence of quadratic spin texture that is unique to altermagnetic materials \cite{Zhu2024-1} and the ferroelectric control of the spin texture is of great interest. Considering the altermagnetic ordering and ferrimagnetic ordering in the $Pc$ and $P2_1$ phases where the $\mathcal{T}$ is absent, we further calculate the band structure and spin texture including SOC effect. Our band calculations show that the CBM is located in $\Gamma$ in both phases ($Pc$ and $P2_1$) with the band splitting around 12 meV and 6 meV, respectively. Interestingly, as is presented in Figs. \ref{Fig5}(c) and (d), PST is discovered around CBM in both phases.

%, whose origin is different from the previous systems with $\mathcal{T}$ \cite{PhysRevLett.97.236601, Tao2018, Lu2020, PhysRevLett.125.216405, PhysRevB.107.035155}.

Given the MPG of $Pc$ and $P2_1$ phases (i.e. $m$ and $2'$), the \textit{\textbf{k $\cdot$ p}} Hamiltonian around CBM in $Pc$ and $P2_1$ phases can be constructed according to the magnetic co-little group of $\Gamma$, that is, $m$ and $2'$. In the $k_x-k_y$ plane in the $Pc$ phase, the Hamiltonian derived from the \textit{\textbf{k $\cdot$ p}} theory around $\Gamma$ with $m_y$ symmetry is:
\begin{equation}
	\begin{split}
		H &= \sigma_x \alpha_1 k_x k_y + \sigma_y (\beta_1 k_x^2 + \beta_2 k_y^2) + \sigma_z \gamma_1 k_x k_y\\		
	\end{split}
\end{equation}
It is obvious that the spin texture around $\Gamma$ is dominated by the quadratic term $\sigma_y (\beta_1 k_x^2 + \beta_2 k_y^2)$, giving rise to the PST along $y$ direction presented in Fig. \ref*{Fig5}(c), which is in accordance with the direction of wFM. According to the Hamiltonian, the direction of $S_x$ will be reversed as the sign of $k_x$ or $k_y$ changes, which is consistent with our DFT calculation. Besides, the component $S_z$ around $\Gamma$ within the region of PST is negligible.

%This indicates the formation of the PST is mainly affected by the wFM, which is consistent with a previous report that the parallel spin component can result in the nonzero wFM macroscopically \cite{PhysRevB.111.054442}. 
 
For the $P2_1$ phase, the Hamiltonian around $\Gamma$ with $\mathcal{T}C_{2y}$ symmetry is:
\begin{equation}
	\begin{split}
		H &= \sigma_x (\alpha_1 k_x^2 + \alpha_2 k_y^2) + \sigma_y \beta_1 k_x k_y \\
		 & + \sigma_z(\gamma_1 k_x^2+\gamma_2 k_y^2)\\
	\end{split}
\end{equation}
Similarly, as shown in Fig. \ref{Fig5}(d),  the quadratic term $\sigma_x (\alpha_1 k_x^2 + \alpha_2 k_y^2)$ dominate the PST along $x$ direction and the direction of $S_y$ will be reversed as the sign of $k_x$ or $k_y$ changes, which is in agreement with our DFT calculation. Also, the component $S_z$ around $\Gamma$ within the region of PST is negligible. Due to the difference of dominated quadratic terms enforced by symmetry in the $Pc$ and $P2_1$ phase, the direction of PST along $y$ can be changed to be $x$ direction through the $Pc$-to-$P2_1$ phase transition by changing OP-HIF.

\section{Conclusion}
In summary, the epitaxial strain engineering with a$\neq$b can be used to create the out-of-plane ferroelectricity in the 2D-RP structure. Besides, in the 2D-RP structure of Ca$_3$Mn$_2$O$_7$, the out-of-plane ferroelectricity can be tuned to favor a transition into a polar phase with only in-plane ferroelectricity, where there is a change of the direction of wFM that is constrained by the epitaxial strain. Our study facilitates discovery of PST allowing for the long spin lifetime in the materials without $\mathcal{T}$, which is in contrast to the previously studied PST in the system with the $\mathcal{T}$ or $\uptau$$\mathcal{T}$ symmetry \cite{Lu2020, PhysRevLett.125.216405, PhysRevLett.125.256603, Lou2020}. Furthermore, a novel approach is proposed to tune the direction of PST by manipulating out-of-plane ferroelectricity in 2D materials.

%Since the time reversal symmetry is absent in the two polar phases, the spin components with the same direction as wFM can take over the effective spin-orbital field comprised of quadratic momentum vectors, leading to a change of PST about a $90^\circ$ in the in-plane by out-of-plane ferroelectricity when the phase transition occurs. Furthermore, a novel approach is proposed to tune the direction of PST by manipulating out-of-plane ferroelectricity in 2D materials.
% Our study facilitates discovery of PST allowing for the long spin lifetime in the materials without $\mathcal{T}$, which is in contrast to the PST in the system with the $\mathcal{T}$ or $\uptau$$\mathcal{T}$ symmetry. And the previously found PST is usually governed by the odd terms of wavevector $k$, such as the linear and cubic terms \cite{Lu2020, PhysRevLett.125.216405, PhysRevLett.125.256603, Lou2020}.

\begin{acknowledgments}
 Y.Z., C.A.J. and X.-Z.L. were supported by the National Natural Science Foundation of China (NSFC) under Grant No. 12474081, the open research fund of Key Laboratory of Quantum Materials and Devices (South-east University), Ministry of Education, the Start-up Research Fund of Southeast University. Y.Z. was supported by the SEU Innovation Capability Enhancement Plan for Doctoral Students (Grant No. CXJH\_SEU 25135). DFT calculations were performed through the high-performance computers, supported by the Big Data Computing Center of Southeast University.
\end{acknowledgments}

%\bibliography{reference}

%\bibliography{reference.bib}% Produces the bibliography via BibTeX.
%\bibliography{reference}
%\bibliographystyle{apsrev4-2}
\bibliography{reference.bib}

\begin{thebibliography}{74}%
\makeatletter
\providecommand \@ifxundefined [1]{%
 \@ifx{#1\undefined}
}%
\providecommand \@ifnum [1]{%
 \ifnum #1\expandafter \@firstoftwo
 \else \expandafter \@secondoftwo
 \fi
}%
\providecommand \@ifx [1]{%
 \ifx #1\expandafter \@firstoftwo
 \else \expandafter \@secondoftwo
 \fi
}%
\providecommand \natexlab [1]{#1}%
\providecommand \enquote  [1]{``#1''}%
\providecommand \bibnamefont  [1]{#1}%
\providecommand \bibfnamefont [1]{#1}%
\providecommand \citenamefont [1]{#1}%
\providecommand \href@noop [0]{\@secondoftwo}%
\providecommand \href [0]{\begingroup \@sanitize@url \@href}%
\providecommand \@href[1]{\@@startlink{#1}\@@href}%
\providecommand \@@href[1]{\endgroup#1\@@endlink}%
\providecommand \@sanitize@url [0]{\catcode `\\12\catcode `\$12\catcode
  `\&12\catcode `\#12\catcode `\^12\catcode `\_12\catcode `\%12\relax}%
\providecommand \@@startlink[1]{}%
\providecommand \@@endlink[0]{}%
\providecommand \url  [0]{\begingroup\@sanitize@url \@url }%
\providecommand \@url [1]{\endgroup\@href {#1}{\urlprefix }}%
\providecommand \urlprefix  [0]{URL }%
\providecommand \Eprint [0]{\href }%
\providecommand \doibase [0]{https://doi.org/}%
\providecommand \selectlanguage [0]{\@gobble}%
\providecommand \bibinfo  [0]{\@secondoftwo}%
\providecommand \bibfield  [0]{\@secondoftwo}%
\providecommand \translation [1]{[#1]}%
\providecommand \BibitemOpen [0]{}%
\providecommand \bibitemStop [0]{}%
\providecommand \bibitemNoStop [0]{.\EOS\space}%
\providecommand \EOS [0]{\spacefactor3000\relax}%
\providecommand \BibitemShut  [1]{\csname bibitem#1\endcsname}%
\let\auto@bib@innerbib\@empty
%</preamble>
\bibitem [{\citenamefont {Chang}\ \emph {et~al.}(2016)\citenamefont {Chang},
  \citenamefont {Liu}, \citenamefont {Lin}, \citenamefont {Wang}, \citenamefont
  {Zhao}, \citenamefont {Zhang}, \citenamefont {Jin}, \citenamefont {Zhong},
  \citenamefont {Hu}, \citenamefont {Duan}, \citenamefont {Zhang},
  \citenamefont {Fu}, \citenamefont {Xue}, \citenamefont {Chen},\ and\
  \citenamefont {Ji}}]{Chang2016}%
  \BibitemOpen
  \bibfield  {author} {\bibinfo {author} {\bibfnamefont {K.}~\bibnamefont
  {Chang}}, \bibinfo {author} {\bibfnamefont {J.}~\bibnamefont {Liu}}, \bibinfo
  {author} {\bibfnamefont {H.}~\bibnamefont {Lin}}, \bibinfo {author}
  {\bibfnamefont {N.}~\bibnamefont {Wang}}, \bibinfo {author} {\bibfnamefont
  {K.}~\bibnamefont {Zhao}}, \bibinfo {author} {\bibfnamefont {A.}~\bibnamefont
  {Zhang}}, \bibinfo {author} {\bibfnamefont {F.}~\bibnamefont {Jin}}, \bibinfo
  {author} {\bibfnamefont {Y.}~\bibnamefont {Zhong}}, \bibinfo {author}
  {\bibfnamefont {X.}~\bibnamefont {Hu}}, \bibinfo {author} {\bibfnamefont
  {W.}~\bibnamefont {Duan}}, \bibinfo {author} {\bibfnamefont {Q.}~\bibnamefont
  {Zhang}}, \bibinfo {author} {\bibfnamefont {L.}~\bibnamefont {Fu}}, \bibinfo
  {author} {\bibfnamefont {Q.-K.}\ \bibnamefont {Xue}}, \bibinfo {author}
  {\bibfnamefont {X.}~\bibnamefont {Chen}},\ and\ \bibinfo {author}
  {\bibfnamefont {S.-H.}\ \bibnamefont {Ji}},\ }\bibfield  {title} {\bibinfo
  {title} {Discovery of robust in-plane ferroelectricity in atomic-thick
  {SnTe}},\ }\href@noop {} {\bibfield  {journal} {\bibinfo  {journal}
  {Science}\ }\textbf {\bibinfo {volume} {353}},\ \bibinfo {pages} {274}
  (\bibinfo {year} {2016})}\BibitemShut {NoStop}%
\bibitem [{\citenamefont {Dong}\ \emph {et~al.}(2019)\citenamefont {Dong},
  \citenamefont {Xiang},\ and\ \citenamefont {Dagotto}}]{Dong2019}%
  \BibitemOpen
  \bibfield  {author} {\bibinfo {author} {\bibfnamefont {S.}~\bibnamefont
  {Dong}}, \bibinfo {author} {\bibfnamefont {H.}~\bibnamefont {Xiang}},\ and\
  \bibinfo {author} {\bibfnamefont {E.}~\bibnamefont {Dagotto}},\ }\bibfield
  {title} {\bibinfo {title} {Magnetoelectricity in multiferroics: a theoretical
  perspective},\ }\href@noop {} {\bibfield  {journal} {\bibinfo  {journal}
  {Natl. Sci. Rev.}\ }\textbf {\bibinfo {volume} {6}},\ \bibinfo {pages} {629}
  (\bibinfo {year} {2019})}\BibitemShut {NoStop}%
\bibitem [{\citenamefont {Banerjee}\ \emph {et~al.}(2022)\citenamefont
  {Banerjee}, \citenamefont {Kashir},\ and\ \citenamefont
  {Kamba}}]{Banerjee2022}%
  \BibitemOpen
  \bibfield  {author} {\bibinfo {author} {\bibfnamefont {W.}~\bibnamefont
  {Banerjee}}, \bibinfo {author} {\bibfnamefont {A.}~\bibnamefont {Kashir}},\
  and\ \bibinfo {author} {\bibfnamefont {S.}~\bibnamefont {Kamba}},\ }\bibfield
   {title} {\bibinfo {title} {Hafnium oxide ({HfO$_2$}) - a multifunctional
  oxide: A review on the prospect and challenges of hafnium oxide in resistive
  switching and ferroelectric memories},\ }\href@noop {} {\bibfield  {journal}
  {\bibinfo  {journal} {Small}\ }\textbf {\bibinfo {volume} {18}},\ \bibinfo
  {pages} {2107575} (\bibinfo {year} {2022})}\BibitemShut {NoStop}%
\bibitem [{\citenamefont {Dawber}\ \emph {et~al.}(2005)\citenamefont {Dawber},
  \citenamefont {Rabe},\ and\ \citenamefont {Scott}}]{Dawber2005}%
  \BibitemOpen
  \bibfield  {author} {\bibinfo {author} {\bibfnamefont {M.}~\bibnamefont
  {Dawber}}, \bibinfo {author} {\bibfnamefont {K.}~\bibnamefont {Rabe}},\ and\
  \bibinfo {author} {\bibfnamefont {J.}~\bibnamefont {Scott}},\ }\bibfield
  {title} {\bibinfo {title} {Physics of thin-film ferroelectric oxides},\
  }\href@noop {} {\bibfield  {journal} {\bibinfo  {journal} {Rev. Mod. Phys.}\
  }\textbf {\bibinfo {volume} {77}},\ \bibinfo {pages} {1083} (\bibinfo {year}
  {2005})}\BibitemShut {NoStop}%
\bibitem [{\citenamefont {Scott}(2007)}]{Scott2007}%
  \BibitemOpen
  \bibfield  {author} {\bibinfo {author} {\bibfnamefont {J.~F.}\ \bibnamefont
  {Scott}},\ }\bibfield  {title} {\bibinfo {title} {Applications of modern
  ferroelectrics},\ }\href@noop {} {\bibfield  {journal} {\bibinfo  {journal}
  {Science}\ }\textbf {\bibinfo {volume} {315}},\ \bibinfo {pages} {954}
  (\bibinfo {year} {2007})}\BibitemShut {NoStop}%
\bibitem [{\citenamefont {Junquera}\ and\ \citenamefont
  {Ghosez}(2003)}]{Junquera2003}%
  \BibitemOpen
  \bibfield  {author} {\bibinfo {author} {\bibfnamefont {J.}~\bibnamefont
  {Junquera}}\ and\ \bibinfo {author} {\bibfnamefont {P.}~\bibnamefont
  {Ghosez}},\ }\bibfield  {title} {\bibinfo {title} {Critical thickness for
  ferroelectricity in perovskite ultrathin films},\ }\href@noop {} {\bibfield
  {journal} {\bibinfo  {journal} {Nature}\ }\textbf {\bibinfo {volume} {422}},\
  \bibinfo {pages} {506} (\bibinfo {year} {2003})}\BibitemShut {NoStop}%
\bibitem [{\citenamefont {Spaldin}(2004)}]{Spaldin2004}%
  \BibitemOpen
  \bibfield  {author} {\bibinfo {author} {\bibfnamefont {N.}~\bibnamefont
  {Spaldin}},\ }\bibfield  {title} {\bibinfo {title} {Fundamental size limits
  in ferroelectricity},\ }\href@noop {} {\bibfield  {journal} {\bibinfo
  {journal} {Science}\ }\textbf {\bibinfo {volume} {304}},\ \bibinfo {pages}
  {1606} (\bibinfo {year} {2004})}\BibitemShut {NoStop}%
\bibitem [{\citenamefont {Song}\ \emph {et~al.}(2022)\citenamefont {Song},
  \citenamefont {Occhialini}, \citenamefont {Ergecen}, \citenamefont {Ilyas},
  \citenamefont {Amoroso}, \citenamefont {Barone}, \citenamefont {Kapeghian},
  \citenamefont {Watanabe}, \citenamefont {Taniguchi}, \citenamefont {Botana},
  \citenamefont {Picozzi}, \citenamefont {Gedik},\ and\ \citenamefont
  {Comin}}]{Song2022}%
  \BibitemOpen
  \bibfield  {author} {\bibinfo {author} {\bibfnamefont {Q.}~\bibnamefont
  {Song}}, \bibinfo {author} {\bibfnamefont {C.~A.}\ \bibnamefont
  {Occhialini}}, \bibinfo {author} {\bibfnamefont {E.}~\bibnamefont {Ergecen}},
  \bibinfo {author} {\bibfnamefont {B.}~\bibnamefont {Ilyas}}, \bibinfo
  {author} {\bibfnamefont {D.}~\bibnamefont {Amoroso}}, \bibinfo {author}
  {\bibfnamefont {P.}~\bibnamefont {Barone}}, \bibinfo {author} {\bibfnamefont
  {J.}~\bibnamefont {Kapeghian}}, \bibinfo {author} {\bibfnamefont
  {K.}~\bibnamefont {Watanabe}}, \bibinfo {author} {\bibfnamefont
  {T.}~\bibnamefont {Taniguchi}}, \bibinfo {author} {\bibfnamefont {A.~S.}\
  \bibnamefont {Botana}}, \bibinfo {author} {\bibfnamefont {S.}~\bibnamefont
  {Picozzi}}, \bibinfo {author} {\bibfnamefont {N.}~\bibnamefont {Gedik}},\
  and\ \bibinfo {author} {\bibfnamefont {R.}~\bibnamefont {Comin}},\ }\bibfield
   {title} {\bibinfo {title} {Evidence for a single-layer van der {Waals}
  multiferroic},\ }\href@noop {} {\bibfield  {journal} {\bibinfo  {journal}
  {Nature}\ }\textbf {\bibinfo {volume} {602}},\ \bibinfo {pages} {601}
  (\bibinfo {year} {2022})}\BibitemShut {NoStop}%
\bibitem [{\citenamefont {Yasuda}\ \emph {et~al.}(2021)\citenamefont {Yasuda},
  \citenamefont {Wang}, \citenamefont {Watanabe}, \citenamefont {Taniguchi},\
  and\ \citenamefont {Jarillo-Herrero}}]{Yasuda2021}%
  \BibitemOpen
  \bibfield  {author} {\bibinfo {author} {\bibfnamefont {K.}~\bibnamefont
  {Yasuda}}, \bibinfo {author} {\bibfnamefont {X.}~\bibnamefont {Wang}},
  \bibinfo {author} {\bibfnamefont {K.}~\bibnamefont {Watanabe}}, \bibinfo
  {author} {\bibfnamefont {T.}~\bibnamefont {Taniguchi}},\ and\ \bibinfo
  {author} {\bibfnamefont {P.}~\bibnamefont {Jarillo-Herrero}},\ }\bibfield
  {title} {\bibinfo {title} {Stacking-engineered ferroelectricity in bilayer
  boron nitride},\ }\href@noop {} {\bibfield  {journal} {\bibinfo  {journal}
  {Science}\ }\textbf {\bibinfo {volume} {372}},\ \bibinfo {pages} {1458}
  (\bibinfo {year} {2021})}\BibitemShut {NoStop}%
\bibitem [{\citenamefont {Liu}\ \emph {et~al.}(2024)\citenamefont {Liu},
  \citenamefont {Ren},\ and\ \citenamefont {Picozzi}}]{PhysRevLett.132.086802}%
  \BibitemOpen
  \bibfield  {author} {\bibinfo {author} {\bibfnamefont {C.}~\bibnamefont
  {Liu}}, \bibinfo {author} {\bibfnamefont {W.}~\bibnamefont {Ren}},\ and\
  \bibinfo {author} {\bibfnamefont {S.}~\bibnamefont {Picozzi}},\ }\bibfield
  {title} {\bibinfo {title} {Spin-chirality-driven multiferroicity in van der
  waals monolayers},\ }\href@noop {} {\bibfield  {journal} {\bibinfo  {journal}
  {Phys. Rev. Lett.}\ }\textbf {\bibinfo {volume} {132}},\ \bibinfo {pages}
  {086802} (\bibinfo {year} {2024})}\BibitemShut {NoStop}%
\bibitem [{\citenamefont {Belianinov}\ \emph {et~al.}(2015)\citenamefont
  {Belianinov}, \citenamefont {He}, \citenamefont {Dziaugys}, \citenamefont
  {Maksymovych}, \citenamefont {Eliseev}, \citenamefont {Borisevich},
  \citenamefont {Morozovska}, \citenamefont {Banys}, \citenamefont
  {Vysochanskii},\ and\ \citenamefont {Kalinin}}]{Belianinov2015}%
  \BibitemOpen
  \bibfield  {author} {\bibinfo {author} {\bibfnamefont {A.}~\bibnamefont
  {Belianinov}}, \bibinfo {author} {\bibfnamefont {Q.}~\bibnamefont {He}},
  \bibinfo {author} {\bibfnamefont {A.}~\bibnamefont {Dziaugys}}, \bibinfo
  {author} {\bibfnamefont {P.}~\bibnamefont {Maksymovych}}, \bibinfo {author}
  {\bibfnamefont {E.}~\bibnamefont {Eliseev}}, \bibinfo {author} {\bibfnamefont
  {A.}~\bibnamefont {Borisevich}}, \bibinfo {author} {\bibfnamefont
  {A.}~\bibnamefont {Morozovska}}, \bibinfo {author} {\bibfnamefont
  {J.}~\bibnamefont {Banys}}, \bibinfo {author} {\bibfnamefont
  {Y.}~\bibnamefont {Vysochanskii}},\ and\ \bibinfo {author} {\bibfnamefont
  {S.~V.}\ \bibnamefont {Kalinin}},\ }\bibfield  {title} {\bibinfo {title}
  {{CuInP$_2$S$_6$} room temperature layered ferroelectric},\ }\href@noop {}
  {\bibfield  {journal} {\bibinfo  {journal} {Nano Lett.}\ }\textbf {\bibinfo
  {volume} {15}},\ \bibinfo {pages} {3808} (\bibinfo {year}
  {2015})}\BibitemShut {NoStop}%
\bibitem [{\citenamefont {Liu}\ \emph {et~al.}(2016)\citenamefont {Liu},
  \citenamefont {You}, \citenamefont {Seyler}, \citenamefont {Li},
  \citenamefont {Yu}, \citenamefont {Lin}, \citenamefont {Wang}, \citenamefont
  {Zhou}, \citenamefont {Wang}, \citenamefont {He}, \citenamefont {Pantelides},
  \citenamefont {Zhou}, \citenamefont {Sharma}, \citenamefont {Xu},
  \citenamefont {Ajayan}, \citenamefont {Wang},\ and\ \citenamefont
  {Liu}}]{Liu2016}%
  \BibitemOpen
  \bibfield  {author} {\bibinfo {author} {\bibfnamefont {F.}~\bibnamefont
  {Liu}}, \bibinfo {author} {\bibfnamefont {L.}~\bibnamefont {You}}, \bibinfo
  {author} {\bibfnamefont {K.~L.}\ \bibnamefont {Seyler}}, \bibinfo {author}
  {\bibfnamefont {X.}~\bibnamefont {Li}}, \bibinfo {author} {\bibfnamefont
  {P.}~\bibnamefont {Yu}}, \bibinfo {author} {\bibfnamefont {J.}~\bibnamefont
  {Lin}}, \bibinfo {author} {\bibfnamefont {X.}~\bibnamefont {Wang}}, \bibinfo
  {author} {\bibfnamefont {J.}~\bibnamefont {Zhou}}, \bibinfo {author}
  {\bibfnamefont {H.}~\bibnamefont {Wang}}, \bibinfo {author} {\bibfnamefont
  {H.}~\bibnamefont {He}}, \bibinfo {author} {\bibfnamefont {S.~T.}\
  \bibnamefont {Pantelides}}, \bibinfo {author} {\bibfnamefont
  {W.}~\bibnamefont {Zhou}}, \bibinfo {author} {\bibfnamefont {P.}~\bibnamefont
  {Sharma}}, \bibinfo {author} {\bibfnamefont {X.}~\bibnamefont {Xu}}, \bibinfo
  {author} {\bibfnamefont {P.~M.}\ \bibnamefont {Ajayan}}, \bibinfo {author}
  {\bibfnamefont {J.}~\bibnamefont {Wang}},\ and\ \bibinfo {author}
  {\bibfnamefont {Z.}~\bibnamefont {Liu}},\ }\bibfield  {title} {\bibinfo
  {title} {Room-temperature ferroelectricity in {CuInP$_2$S$_6$} ultrathin
  flakes},\ }\href@noop {} {\bibfield  {journal} {\bibinfo  {journal} {Nat.
  Commun.}\ }\textbf {\bibinfo {volume} {7}},\ \bibinfo {pages} {12357}
  (\bibinfo {year} {2016})}\BibitemShut {NoStop}%
\bibitem [{\citenamefont {Li}\ and\ \citenamefont {Wu}(2017)}]{Li2017}%
  \BibitemOpen
  \bibfield  {author} {\bibinfo {author} {\bibfnamefont {L.}~\bibnamefont
  {Li}}\ and\ \bibinfo {author} {\bibfnamefont {M.}~\bibnamefont {Wu}},\
  }\bibfield  {title} {\bibinfo {title} {Binary compound bilayer and multilayer
  with vertical polarizations: Two-dimensional ferroelectrics, multiferroics,
  and nanogenerators},\ }\href@noop {} {\bibfield  {journal} {\bibinfo
  {journal} {Acs Nano}\ }\textbf {\bibinfo {volume} {11}},\ \bibinfo {pages}
  {6382} (\bibinfo {year} {2017})}\BibitemShut {NoStop}%
\bibitem [{\citenamefont {Fei}\ \emph {et~al.}(2018)\citenamefont {Fei},
  \citenamefont {Zhao}, \citenamefont {Palomaki}, \citenamefont {Sun},
  \citenamefont {Miller}, \citenamefont {Zhao}, \citenamefont {Yan},
  \citenamefont {Xu},\ and\ \citenamefont {Cobden}}]{Fei2018}%
  \BibitemOpen
  \bibfield  {author} {\bibinfo {author} {\bibfnamefont {Z.}~\bibnamefont
  {Fei}}, \bibinfo {author} {\bibfnamefont {W.}~\bibnamefont {Zhao}}, \bibinfo
  {author} {\bibfnamefont {T.~A.}\ \bibnamefont {Palomaki}}, \bibinfo {author}
  {\bibfnamefont {B.}~\bibnamefont {Sun}}, \bibinfo {author} {\bibfnamefont
  {M.~K.}\ \bibnamefont {Miller}}, \bibinfo {author} {\bibfnamefont
  {Z.}~\bibnamefont {Zhao}}, \bibinfo {author} {\bibfnamefont {J.}~\bibnamefont
  {Yan}}, \bibinfo {author} {\bibfnamefont {X.}~\bibnamefont {Xu}},\ and\
  \bibinfo {author} {\bibfnamefont {D.~H.}\ \bibnamefont {Cobden}},\ }\bibfield
   {title} {\bibinfo {title} {Ferroelectric switching of a two-dimensional
  metal},\ }\href@noop {} {\bibfield  {journal} {\bibinfo  {journal} {Nature}\
  }\textbf {\bibinfo {volume} {560}},\ \bibinfo {pages} {336} (\bibinfo {year}
  {2018})}\BibitemShut {NoStop}%
\bibitem [{\citenamefont {Ding}\ \emph {et~al.}(2017)\citenamefont {Ding},
  \citenamefont {Zhu}, \citenamefont {Wang}, \citenamefont {Gao}, \citenamefont
  {Xiao}, \citenamefont {Gu}, \citenamefont {Zhang},\ and\ \citenamefont
  {Zhu}}]{Ding2017}%
  \BibitemOpen
  \bibfield  {author} {\bibinfo {author} {\bibfnamefont {W.}~\bibnamefont
  {Ding}}, \bibinfo {author} {\bibfnamefont {J.}~\bibnamefont {Zhu}}, \bibinfo
  {author} {\bibfnamefont {Z.}~\bibnamefont {Wang}}, \bibinfo {author}
  {\bibfnamefont {Y.}~\bibnamefont {Gao}}, \bibinfo {author} {\bibfnamefont
  {D.}~\bibnamefont {Xiao}}, \bibinfo {author} {\bibfnamefont {Y.}~\bibnamefont
  {Gu}}, \bibinfo {author} {\bibfnamefont {Z.}~\bibnamefont {Zhang}},\ and\
  \bibinfo {author} {\bibfnamefont {W.}~\bibnamefont {Zhu}},\ }\bibfield
  {title} {\bibinfo {title} {Prediction of intrinsic two-dimensional
  ferroelectrics in {In$_2$Se$_3$} and other {III}$_2$-{VI}$_3$ van der waals
  materials},\ }\href@noop {} {\bibfield  {journal} {\bibinfo  {journal} {Nat.
  Commun.}\ }\textbf {\bibinfo {volume} {8}},\ \bibinfo {pages} {14956}
  (\bibinfo {year} {2017})}\BibitemShut {NoStop}%
\bibitem [{\citenamefont {Zhou}\ \emph {et~al.}(2017)\citenamefont {Zhou},
  \citenamefont {Wu}, \citenamefont {Zhu}, \citenamefont {Cho}, \citenamefont
  {He}, \citenamefont {Yang}, \citenamefont {Herrera}, \citenamefont {Chu},
  \citenamefont {Han}, \citenamefont {Downer}, \citenamefont {Peng},\ and\
  \citenamefont {Lai}}]{Zhou2017}%
  \BibitemOpen
  \bibfield  {author} {\bibinfo {author} {\bibfnamefont {Y.}~\bibnamefont
  {Zhou}}, \bibinfo {author} {\bibfnamefont {D.}~\bibnamefont {Wu}}, \bibinfo
  {author} {\bibfnamefont {Y.}~\bibnamefont {Zhu}}, \bibinfo {author}
  {\bibfnamefont {Y.}~\bibnamefont {Cho}}, \bibinfo {author} {\bibfnamefont
  {Q.}~\bibnamefont {He}}, \bibinfo {author} {\bibfnamefont {X.}~\bibnamefont
  {Yang}}, \bibinfo {author} {\bibfnamefont {K.}~\bibnamefont {Herrera}},
  \bibinfo {author} {\bibfnamefont {Z.}~\bibnamefont {Chu}}, \bibinfo {author}
  {\bibfnamefont {Y.}~\bibnamefont {Han}}, \bibinfo {author} {\bibfnamefont
  {M.~C.}\ \bibnamefont {Downer}}, \bibinfo {author} {\bibfnamefont
  {H.}~\bibnamefont {Peng}},\ and\ \bibinfo {author} {\bibfnamefont
  {K.}~\bibnamefont {Lai}},\ }\bibfield  {title} {\bibinfo {title}
  {Out-of-plane piezoelectricity and ferroelectricity in layered
  $\alpha$-{In$_2$Se$_3$} nanoflakes},\ }\href@noop {} {\bibfield  {journal}
  {\bibinfo  {journal} {Nano Lett.}\ }\textbf {\bibinfo {volume} {17}},\
  \bibinfo {pages} {5508} (\bibinfo {year} {2017})}\BibitemShut {NoStop}%
\bibitem [{\citenamefont {Hu}\ and\ \citenamefont {Kan}(2019)}]{Hu2019}%
  \BibitemOpen
  \bibfield  {author} {\bibinfo {author} {\bibfnamefont {T.}~\bibnamefont
  {Hu}}\ and\ \bibinfo {author} {\bibfnamefont {E.}~\bibnamefont {Kan}},\
  }\bibfield  {title} {\bibinfo {title} {Progress and prospects in
  low-dimensional multiferroic materials},\ }\href@noop {} {\bibfield
  {journal} {\bibinfo  {journal} {Wires. Comput. Mol. Sci.}\ }\textbf {\bibinfo
  {volume} {9}},\ \bibinfo {pages} {e1409} (\bibinfo {year}
  {2019})}\BibitemShut {NoStop}%
\bibitem [{\citenamefont {Spaldin}\ and\ \citenamefont
  {Ramesh}(2019)}]{Spaldin2019}%
  \BibitemOpen
  \bibfield  {author} {\bibinfo {author} {\bibfnamefont {N.~A.}\ \bibnamefont
  {Spaldin}}\ and\ \bibinfo {author} {\bibfnamefont {R.}~\bibnamefont
  {Ramesh}},\ }\bibfield  {title} {\bibinfo {title} {Advances in
  magnetoelectric multiferroics},\ }\href@noop {} {\bibfield  {journal}
  {\bibinfo  {journal} {Nat. Mater.}\ }\textbf {\bibinfo {volume} {18}},\
  \bibinfo {pages} {203} (\bibinfo {year} {2019})}\BibitemShut {NoStop}%
\bibitem [{\citenamefont {Spaldin}(2017)}]{Spaldin2017}%
  \BibitemOpen
  \bibfield  {author} {\bibinfo {author} {\bibfnamefont {N.~A.}\ \bibnamefont
  {Spaldin}},\ }\bibfield  {title} {\bibinfo {title} {Multiferroics: Past,
  present, and future},\ }\href@noop {} {\bibfield  {journal} {\bibinfo
  {journal} {Mrs Bull.}\ }\textbf {\bibinfo {volume} {42}},\ \bibinfo {pages}
  {385} (\bibinfo {year} {2017})}\BibitemShut {NoStop}%
\bibitem [{\citenamefont {Dong}\ \emph {et~al.}(2015)\citenamefont {Dong},
  \citenamefont {Liu}, \citenamefont {Cheong},\ and\ \citenamefont
  {Ren}}]{Dong2015}%
  \BibitemOpen
  \bibfield  {author} {\bibinfo {author} {\bibfnamefont {S.}~\bibnamefont
  {Dong}}, \bibinfo {author} {\bibfnamefont {J.-M.}\ \bibnamefont {Liu}},
  \bibinfo {author} {\bibfnamefont {S.-W.}\ \bibnamefont {Cheong}},\ and\
  \bibinfo {author} {\bibfnamefont {Z.}~\bibnamefont {Ren}},\ }\bibfield
  {title} {\bibinfo {title} {Multiferroic materials and magnetoelectric
  physics: symmetry, entanglement, excitation, and topology},\ }\href@noop {}
  {\bibfield  {journal} {\bibinfo  {journal} {Adv. Phys.}\ }\textbf {\bibinfo
  {volume} {64}},\ \bibinfo {pages} {519} (\bibinfo {year} {2015})}\BibitemShut
  {NoStop}%
\bibitem [{\citenamefont {Nicolosi}\ \emph {et~al.}(2013)\citenamefont
  {Nicolosi}, \citenamefont {Chhowalla}, \citenamefont {Kanatzidis},
  \citenamefont {Strano},\ and\ \citenamefont {Coleman}}]{Nicolosi2013}%
  \BibitemOpen
  \bibfield  {author} {\bibinfo {author} {\bibfnamefont {V.}~\bibnamefont
  {Nicolosi}}, \bibinfo {author} {\bibfnamefont {M.}~\bibnamefont {Chhowalla}},
  \bibinfo {author} {\bibfnamefont {M.~G.}\ \bibnamefont {Kanatzidis}},
  \bibinfo {author} {\bibfnamefont {M.~S.}\ \bibnamefont {Strano}},\ and\
  \bibinfo {author} {\bibfnamefont {J.~N.}\ \bibnamefont {Coleman}},\
  }\bibfield  {title} {\bibinfo {title} {Liquid exfoliation of layered
  materials},\ }\href@noop {} {\bibfield  {journal} {\bibinfo  {journal}
  {Science}\ }\textbf {\bibinfo {volume} {340}},\ \bibinfo {pages} {1420}
  (\bibinfo {year} {2013})}\BibitemShut {NoStop}%
\bibitem [{\citenamefont {Coleman}\ \emph {et~al.}(2011)\citenamefont
  {Coleman}, \citenamefont {Lotya}, \citenamefont {O'Neill}, \citenamefont
  {Bergin}, \citenamefont {King}, \citenamefont {Khan}, \citenamefont {Young},
  \citenamefont {Gaucher}, \citenamefont {De}, \citenamefont {Smith},
  \citenamefont {Shvets}, \citenamefont {Arora}, \citenamefont {Stanton},
  \citenamefont {Kim}, \citenamefont {Lee}, \citenamefont {Kim}, \citenamefont
  {Duesberg}, \citenamefont {Hallam}, \citenamefont {Boland}, \citenamefont
  {Wang}, \citenamefont {Donegan}, \citenamefont {Grunlan}, \citenamefont
  {Moriarty}, \citenamefont {Shmeliov}, \citenamefont {Nicholls}, \citenamefont
  {Perkins}, \citenamefont {Grieveson}, \citenamefont {Theuwissen},
  \citenamefont {McComb}, \citenamefont {Nellist},\ and\ \citenamefont
  {Nicolosi}}]{Coleman2011}%
  \BibitemOpen
  \bibfield  {author} {\bibinfo {author} {\bibfnamefont {J.~N.}\ \bibnamefont
  {Coleman}}, \bibinfo {author} {\bibfnamefont {M.}~\bibnamefont {Lotya}},
  \bibinfo {author} {\bibfnamefont {A.}~\bibnamefont {O'Neill}}, \bibinfo
  {author} {\bibfnamefont {S.~D.}\ \bibnamefont {Bergin}}, \bibinfo {author}
  {\bibfnamefont {P.~J.}\ \bibnamefont {King}}, \bibinfo {author}
  {\bibfnamefont {U.}~\bibnamefont {Khan}}, \bibinfo {author} {\bibfnamefont
  {K.}~\bibnamefont {Young}}, \bibinfo {author} {\bibfnamefont
  {A.}~\bibnamefont {Gaucher}}, \bibinfo {author} {\bibfnamefont
  {S.}~\bibnamefont {De}}, \bibinfo {author} {\bibfnamefont {R.~J.}\
  \bibnamefont {Smith}}, \bibinfo {author} {\bibfnamefont {I.~V.}\ \bibnamefont
  {Shvets}}, \bibinfo {author} {\bibfnamefont {S.~K.}\ \bibnamefont {Arora}},
  \bibinfo {author} {\bibfnamefont {G.}~\bibnamefont {Stanton}}, \bibinfo
  {author} {\bibfnamefont {H.-Y.}\ \bibnamefont {Kim}}, \bibinfo {author}
  {\bibfnamefont {K.}~\bibnamefont {Lee}}, \bibinfo {author} {\bibfnamefont
  {G.~T.}\ \bibnamefont {Kim}}, \bibinfo {author} {\bibfnamefont {G.~S.}\
  \bibnamefont {Duesberg}}, \bibinfo {author} {\bibfnamefont {T.}~\bibnamefont
  {Hallam}}, \bibinfo {author} {\bibfnamefont {J.~J.}\ \bibnamefont {Boland}},
  \bibinfo {author} {\bibfnamefont {J.~J.}\ \bibnamefont {Wang}}, \bibinfo
  {author} {\bibfnamefont {J.~F.}\ \bibnamefont {Donegan}}, \bibinfo {author}
  {\bibfnamefont {J.~C.}\ \bibnamefont {Grunlan}}, \bibinfo {author}
  {\bibfnamefont {G.}~\bibnamefont {Moriarty}}, \bibinfo {author}
  {\bibfnamefont {A.}~\bibnamefont {Shmeliov}}, \bibinfo {author}
  {\bibfnamefont {R.~J.}\ \bibnamefont {Nicholls}}, \bibinfo {author}
  {\bibfnamefont {J.~M.}\ \bibnamefont {Perkins}}, \bibinfo {author}
  {\bibfnamefont {E.~M.}\ \bibnamefont {Grieveson}}, \bibinfo {author}
  {\bibfnamefont {K.}~\bibnamefont {Theuwissen}}, \bibinfo {author}
  {\bibfnamefont {D.~W.}\ \bibnamefont {McComb}}, \bibinfo {author}
  {\bibfnamefont {P.~D.}\ \bibnamefont {Nellist}},\ and\ \bibinfo {author}
  {\bibfnamefont {V.}~\bibnamefont {Nicolosi}},\ }\bibfield  {title} {\bibinfo
  {title} {Two-dimensional nanosheets produced by liquid exfoliation of layered
  materials},\ }\href@noop {} {\bibfield  {journal} {\bibinfo  {journal}
  {Science}\ }\textbf {\bibinfo {volume} {331}},\ \bibinfo {pages} {568}
  (\bibinfo {year} {2011})}\BibitemShut {NoStop}%
\bibitem [{\citenamefont {Balan}\ \emph {et~al.}(2018)\citenamefont {Balan},
  \citenamefont {Radhakrishnan}, \citenamefont {Woellner}, \citenamefont
  {Sinha}, \citenamefont {Deng}, \citenamefont {de~los Reyes}, \citenamefont
  {Rao}, \citenamefont {Paulose}, \citenamefont {Neupane}, \citenamefont
  {Apte}, \citenamefont {Kochat}, \citenamefont {Vajtai}, \citenamefont
  {Harutyunyan}, \citenamefont {Chu}, \citenamefont {Costin}, \citenamefont
  {Galvao}, \citenamefont {Marti}, \citenamefont {van Aken}, \citenamefont
  {Varghese}, \citenamefont {Tiwary}, \citenamefont {Iyer},\ and\ \citenamefont
  {Ajayan}}]{Balan2018}%
  \BibitemOpen
  \bibfield  {author} {\bibinfo {author} {\bibfnamefont {A.~P.}\ \bibnamefont
  {Balan}}, \bibinfo {author} {\bibfnamefont {S.}~\bibnamefont
  {Radhakrishnan}}, \bibinfo {author} {\bibfnamefont {C.~F.}\ \bibnamefont
  {Woellner}}, \bibinfo {author} {\bibfnamefont {S.~K.}\ \bibnamefont {Sinha}},
  \bibinfo {author} {\bibfnamefont {L.}~\bibnamefont {Deng}}, \bibinfo {author}
  {\bibfnamefont {C.}~\bibnamefont {de~los Reyes}}, \bibinfo {author}
  {\bibfnamefont {B.~M.}\ \bibnamefont {Rao}}, \bibinfo {author} {\bibfnamefont
  {M.}~\bibnamefont {Paulose}}, \bibinfo {author} {\bibfnamefont
  {R.}~\bibnamefont {Neupane}}, \bibinfo {author} {\bibfnamefont
  {A.}~\bibnamefont {Apte}}, \bibinfo {author} {\bibfnamefont {V.}~\bibnamefont
  {Kochat}}, \bibinfo {author} {\bibfnamefont {R.}~\bibnamefont {Vajtai}},
  \bibinfo {author} {\bibfnamefont {A.~R.}\ \bibnamefont {Harutyunyan}},
  \bibinfo {author} {\bibfnamefont {C.-W.}\ \bibnamefont {Chu}}, \bibinfo
  {author} {\bibfnamefont {G.}~\bibnamefont {Costin}}, \bibinfo {author}
  {\bibfnamefont {D.~S.}\ \bibnamefont {Galvao}}, \bibinfo {author}
  {\bibfnamefont {A.~A.}\ \bibnamefont {Marti}}, \bibinfo {author}
  {\bibfnamefont {P.~A.}\ \bibnamefont {van Aken}}, \bibinfo {author}
  {\bibfnamefont {O.~K.}\ \bibnamefont {Varghese}}, \bibinfo {author}
  {\bibfnamefont {C.~S.}\ \bibnamefont {Tiwary}}, \bibinfo {author}
  {\bibfnamefont {A.~M. M.~R.}\ \bibnamefont {Iyer}},\ and\ \bibinfo {author}
  {\bibfnamefont {P.~M.}\ \bibnamefont {Ajayan}},\ }\bibfield  {title}
  {\bibinfo {title} {Exfoliation of a non-van der waals material from iron ore
  hematite},\ }\href@noop {} {\bibfield  {journal} {\bibinfo  {journal} {Nat.
  Nanotechnol.}\ }\textbf {\bibinfo {volume} {13}},\ \bibinfo {pages} {602}
  (\bibinfo {year} {2018})}\BibitemShut {NoStop}%
\bibitem [{\citenamefont {Ji}\ \emph {et~al.}(2019)\citenamefont {Ji},
  \citenamefont {Cai}, \citenamefont {Paudel}, \citenamefont {Sun},
  \citenamefont {Zhang}, \citenamefont {Han}, \citenamefont {Wei},
  \citenamefont {Zang}, \citenamefont {Gu}, \citenamefont {Zhang},
  \citenamefont {Gao}, \citenamefont {Huyan}, \citenamefont {Guo},
  \citenamefont {Wu}, \citenamefont {Gu}, \citenamefont {Tsymbal},
  \citenamefont {Wang}, \citenamefont {Nie},\ and\ \citenamefont
  {Pan}}]{Ji2019}%
  \BibitemOpen
  \bibfield  {author} {\bibinfo {author} {\bibfnamefont {D.}~\bibnamefont
  {Ji}}, \bibinfo {author} {\bibfnamefont {S.}~\bibnamefont {Cai}}, \bibinfo
  {author} {\bibfnamefont {T.~R.}\ \bibnamefont {Paudel}}, \bibinfo {author}
  {\bibfnamefont {H.}~\bibnamefont {Sun}}, \bibinfo {author} {\bibfnamefont
  {C.}~\bibnamefont {Zhang}}, \bibinfo {author} {\bibfnamefont
  {L.}~\bibnamefont {Han}}, \bibinfo {author} {\bibfnamefont {Y.}~\bibnamefont
  {Wei}}, \bibinfo {author} {\bibfnamefont {Y.}~\bibnamefont {Zang}}, \bibinfo
  {author} {\bibfnamefont {M.}~\bibnamefont {Gu}}, \bibinfo {author}
  {\bibfnamefont {Y.}~\bibnamefont {Zhang}}, \bibinfo {author} {\bibfnamefont
  {W.}~\bibnamefont {Gao}}, \bibinfo {author} {\bibfnamefont {H.}~\bibnamefont
  {Huyan}}, \bibinfo {author} {\bibfnamefont {W.}~\bibnamefont {Guo}}, \bibinfo
  {author} {\bibfnamefont {D.}~\bibnamefont {Wu}}, \bibinfo {author}
  {\bibfnamefont {Z.}~\bibnamefont {Gu}}, \bibinfo {author} {\bibfnamefont
  {E.~Y.}\ \bibnamefont {Tsymbal}}, \bibinfo {author} {\bibfnamefont
  {P.}~\bibnamefont {Wang}}, \bibinfo {author} {\bibfnamefont {Y.}~\bibnamefont
  {Nie}},\ and\ \bibinfo {author} {\bibfnamefont {X.}~\bibnamefont {Pan}},\
  }\bibfield  {title} {\bibinfo {title} {Freestanding crystalline oxide
  perovskites down to the monolayer limit},\ }\href@noop {} {\bibfield
  {journal} {\bibinfo  {journal} {Nature}\ }\textbf {\bibinfo {volume} {570}},\
  \bibinfo {pages} {87} (\bibinfo {year} {2019})}\BibitemShut {NoStop}%
\bibitem [{\citenamefont {Kresse}\ and\ \citenamefont
  {Furthmuller}(1996{\natexlab{a}})}]{Kresse1996}%
  \BibitemOpen
  \bibfield  {author} {\bibinfo {author} {\bibfnamefont {G.}~\bibnamefont
  {Kresse}}\ and\ \bibinfo {author} {\bibfnamefont {J.}~\bibnamefont
  {Furthmuller}},\ }\bibfield  {title} {\bibinfo {title} {Efficient iterative
  schemes for ab initio total-energy calculations using a plane-wave basis
  set},\ }\href@noop {} {\bibfield  {journal} {\bibinfo  {journal} {Phys. Rev.
  B}\ }\textbf {\bibinfo {volume} {54}},\ \bibinfo {pages} {11169} (\bibinfo
  {year} {1996}{\natexlab{a}})}\BibitemShut {NoStop}%
\bibitem [{\citenamefont {Kresse}\ and\ \citenamefont
  {Furthmuller}(1996{\natexlab{b}})}]{Kresse19961}%
  \BibitemOpen
  \bibfield  {author} {\bibinfo {author} {\bibfnamefont {G.}~\bibnamefont
  {Kresse}}\ and\ \bibinfo {author} {\bibfnamefont {J.}~\bibnamefont
  {Furthmuller}},\ }\bibfield  {title} {\bibinfo {title} {Efficiency of
  ab-initio total energy calculations for metals and semiconductors using a
  plane-wave basis set},\ }\href@noop {} {\bibfield  {journal} {\bibinfo
  {journal} {Computat. Mater. Sci.}\ }\textbf {\bibinfo {volume} {6}},\
  \bibinfo {pages} {15} (\bibinfo {year} {1996}{\natexlab{b}})}\BibitemShut
  {NoStop}%
\bibitem [{\citenamefont {Blochl}(1994)}]{BLOCHL1994}%
  \BibitemOpen
  \bibfield  {author} {\bibinfo {author} {\bibfnamefont {P.}~\bibnamefont
  {Blochl}},\ }\bibfield  {title} {\bibinfo {title} {Projector augmented-wave
  method},\ }\href@noop {} {\bibfield  {journal} {\bibinfo  {journal} {Phys.
  Rev. B}\ }\textbf {\bibinfo {volume} {50}},\ \bibinfo {pages} {17953}
  (\bibinfo {year} {1994})}\BibitemShut {NoStop}%
\bibitem [{\citenamefont {Kresse}\ and\ \citenamefont
  {Joubert}(1999)}]{Kresse1999}%
  \BibitemOpen
  \bibfield  {author} {\bibinfo {author} {\bibfnamefont {G.}~\bibnamefont
  {Kresse}}\ and\ \bibinfo {author} {\bibfnamefont {D.}~\bibnamefont
  {Joubert}},\ }\bibfield  {title} {\bibinfo {title} {From ultrasoft
  pseudopotentials to the projector augmented-wave method},\ }\href@noop {}
  {\bibfield  {journal} {\bibinfo  {journal} {Phys. Rev. B}\ }\textbf {\bibinfo
  {volume} {59}},\ \bibinfo {pages} {1758} (\bibinfo {year}
  {1999})}\BibitemShut {NoStop}%
\bibitem [{\citenamefont {Wang}\ and\ \citenamefont {Perdew}(1991)}]{WANG1991}%
  \BibitemOpen
  \bibfield  {author} {\bibinfo {author} {\bibfnamefont {Y.}~\bibnamefont
  {Wang}}\ and\ \bibinfo {author} {\bibfnamefont {J.}~\bibnamefont {Perdew}},\
  }\bibfield  {title} {\bibinfo {title} {Correlation hole of the spin-polarized
  electron-gas, with exact small-wave-vector and high-density scaling},\
  }\href@noop {} {\bibfield  {journal} {\bibinfo  {journal} {Phys. Rev. B}\
  }\textbf {\bibinfo {volume} {44}},\ \bibinfo {pages} {13298} (\bibinfo {year}
  {1991})}\BibitemShut {NoStop}%
\bibitem [{\citenamefont {Dudarev}\ \emph {et~al.}(1998)\citenamefont
  {Dudarev}, \citenamefont {Botton}, \citenamefont {Savrasov}, \citenamefont
  {Humphreys},\ and\ \citenamefont {Sutton}}]{Dudarev1998}%
  \BibitemOpen
  \bibfield  {author} {\bibinfo {author} {\bibfnamefont {S.}~\bibnamefont
  {Dudarev}}, \bibinfo {author} {\bibfnamefont {G.}~\bibnamefont {Botton}},
  \bibinfo {author} {\bibfnamefont {S.}~\bibnamefont {Savrasov}}, \bibinfo
  {author} {\bibfnamefont {C.}~\bibnamefont {Humphreys}},\ and\ \bibinfo
  {author} {\bibfnamefont {A.}~\bibnamefont {Sutton}},\ }\bibfield  {title}
  {\bibinfo {title} {Electron-energy-loss spectra and the structural stability
  of nickel oxide: {An LSDA+U} study},\ }\href@noop {} {\bibfield  {journal}
  {\bibinfo  {journal} {Phys. Rev. B}\ }\textbf {\bibinfo {volume} {57}},\
  \bibinfo {pages} {1505} (\bibinfo {year} {1998})}\BibitemShut {NoStop}%
\bibitem [{\citenamefont {Togo}\ and\ \citenamefont {Tanaka}(2015)}]{Togo2015}%
  \BibitemOpen
  \bibfield  {author} {\bibinfo {author} {\bibfnamefont {A.}~\bibnamefont
  {Togo}}\ and\ \bibinfo {author} {\bibfnamefont {I.}~\bibnamefont {Tanaka}},\
  }\bibfield  {title} {\bibinfo {title} {First principles phonon calculations
  in materials science},\ }\href@noop {} {\bibfield  {journal} {\bibinfo
  {journal} {Scripta Mater.}\ }\textbf {\bibinfo {volume} {108}},\ \bibinfo
  {pages} {1} (\bibinfo {year} {2015})}\BibitemShut {NoStop}%
\bibitem [{\citenamefont {Kingsmith}\ and\ \citenamefont
  {Vanderbilt}(1993)}]{KINGSMITH1993}%
  \BibitemOpen
  \bibfield  {author} {\bibinfo {author} {\bibfnamefont {R.}~\bibnamefont
  {Kingsmith}}\ and\ \bibinfo {author} {\bibfnamefont {D.}~\bibnamefont
  {Vanderbilt}},\ }\bibfield  {title} {\bibinfo {title} {Theory of polarization
  of crystalline solids},\ }\href@noop {} {\bibfield  {journal} {\bibinfo
  {journal} {Phys. Rev. B}\ }\textbf {\bibinfo {volume} {47}},\ \bibinfo
  {pages} {1651} (\bibinfo {year} {1993})}\BibitemShut {NoStop}%
\bibitem [{\citenamefont {Sheppard}\ \emph {et~al.}(2012)\citenamefont
  {Sheppard}, \citenamefont {Xiao}, \citenamefont {Chemelewski}, \citenamefont
  {Johnson},\ and\ \citenamefont {Henkelman}}]{Sheppard2012}%
  \BibitemOpen
  \bibfield  {author} {\bibinfo {author} {\bibfnamefont {D.}~\bibnamefont
  {Sheppard}}, \bibinfo {author} {\bibfnamefont {P.}~\bibnamefont {Xiao}},
  \bibinfo {author} {\bibfnamefont {W.}~\bibnamefont {Chemelewski}}, \bibinfo
  {author} {\bibfnamefont {D.~D.}\ \bibnamefont {Johnson}},\ and\ \bibinfo
  {author} {\bibfnamefont {G.}~\bibnamefont {Henkelman}},\ }\bibfield  {title}
  {\bibinfo {title} {A generalized solid-state nudged elastic band method},\
  }\href@noop {} {\bibfield  {journal} {\bibinfo  {journal} {J. Chem. Phys.}\
  }\textbf {\bibinfo {volume} {136}},\ \bibinfo {pages} {074103} (\bibinfo
  {year} {2012})}\BibitemShut {NoStop}%
\bibitem [{\citenamefont {Campbell}\ \emph {et~al.}(2006)\citenamefont
  {Campbell}, \citenamefont {Stokes}, \citenamefont {Tanner},\ and\
  \citenamefont {Hatch}}]{Campbell2006}%
  \BibitemOpen
  \bibfield  {author} {\bibinfo {author} {\bibfnamefont {B.~J.}\ \bibnamefont
  {Campbell}}, \bibinfo {author} {\bibfnamefont {H.~T.}\ \bibnamefont
  {Stokes}}, \bibinfo {author} {\bibfnamefont {D.~E.}\ \bibnamefont {Tanner}},\
  and\ \bibinfo {author} {\bibfnamefont {D.~M.}\ \bibnamefont {Hatch}},\
  }\bibfield  {title} {\bibinfo {title} {$isodisplace$:: a web-based tool for
  exploring structural distortions},\ }\href@noop {} {\bibfield  {journal}
  {\bibinfo  {journal} {J. Appl. Cryst.}\ }\textbf {\bibinfo {volume} {39}},\
  \bibinfo {pages} {607} (\bibinfo {year} {2006})}\BibitemShut {NoStop}%
\bibitem [{\citenamefont {Benedek}\ and\ \citenamefont
  {Fennie}(2011{\natexlab{a}})}]{PhysRevLett.106.107204}%
  \BibitemOpen
  \bibfield  {author} {\bibinfo {author} {\bibfnamefont {N.~A.}\ \bibnamefont
  {Benedek}}\ and\ \bibinfo {author} {\bibfnamefont {C.~J.}\ \bibnamefont
  {Fennie}},\ }\bibfield  {title} {\bibinfo {title} {Hybrid improper
  ferroelectricity: A mechanism for controllable polarization-magnetization
  coupling},\ }\href@noop {} {\bibfield  {journal} {\bibinfo  {journal} {Phys.
  Rev. Lett.}\ }\textbf {\bibinfo {volume} {106}},\ \bibinfo {pages} {107204}
  (\bibinfo {year} {2011}{\natexlab{a}})}\BibitemShut {NoStop}%
\bibitem [{\citenamefont {Chu}\ \emph {et~al.}(2008)\citenamefont {Chu},
  \citenamefont {Martin}, \citenamefont {Holcomb}, \citenamefont {Gajek},
  \citenamefont {Han}, \citenamefont {He}, \citenamefont {Balke}, \citenamefont
  {Yang}, \citenamefont {Lee}, \citenamefont {Hu}, \citenamefont {Zhan},
  \citenamefont {Yang}, \citenamefont {Fraile-Rodriguez}, \citenamefont
  {Scholl}, \citenamefont {Wang},\ and\ \citenamefont
  {Ramesh}}]{ChuYing-Hao2008}%
  \BibitemOpen
  \bibfield  {author} {\bibinfo {author} {\bibfnamefont {Y.-H.}\ \bibnamefont
  {Chu}}, \bibinfo {author} {\bibfnamefont {L.~W.}\ \bibnamefont {Martin}},
  \bibinfo {author} {\bibfnamefont {M.~B.}\ \bibnamefont {Holcomb}}, \bibinfo
  {author} {\bibfnamefont {M.}~\bibnamefont {Gajek}}, \bibinfo {author}
  {\bibfnamefont {S.-J.}\ \bibnamefont {Han}}, \bibinfo {author} {\bibfnamefont
  {Q.}~\bibnamefont {He}}, \bibinfo {author} {\bibfnamefont {N.}~\bibnamefont
  {Balke}}, \bibinfo {author} {\bibfnamefont {C.-H.}\ \bibnamefont {Yang}},
  \bibinfo {author} {\bibfnamefont {D.}~\bibnamefont {Lee}}, \bibinfo {author}
  {\bibfnamefont {W.}~\bibnamefont {Hu}}, \bibinfo {author} {\bibfnamefont
  {Q.}~\bibnamefont {Zhan}}, \bibinfo {author} {\bibfnamefont {P.-L.}\
  \bibnamefont {Yang}}, \bibinfo {author} {\bibfnamefont {A.}~\bibnamefont
  {Fraile-Rodriguez}}, \bibinfo {author} {\bibfnamefont {A.}~\bibnamefont
  {Scholl}}, \bibinfo {author} {\bibfnamefont {S.~X.}\ \bibnamefont {Wang}},\
  and\ \bibinfo {author} {\bibfnamefont {R.}~\bibnamefont {Ramesh}},\
  }\bibfield  {title} {\bibinfo {title} {Electric-field control of local
  ferromagnetism using a magnetoelectric multiferroic},\ }\href@noop {}
  {\bibfield  {journal} {\bibinfo  {journal} {Nat. Mater.}\ }\textbf {\bibinfo
  {volume} {7}},\ \bibinfo {pages} {478} (\bibinfo {year} {2008})}\BibitemShut
  {NoStop}%
\bibitem [{\citenamefont {Zhao}\ \emph {et~al.}(2006)\citenamefont {Zhao},
  \citenamefont {Scholl}, \citenamefont {Zavaliche}, \citenamefont {Lee},
  \citenamefont {Barry}, \citenamefont {Doran}, \citenamefont {Cruz},
  \citenamefont {Chu}, \citenamefont {Ederer}, \citenamefont {Spaldin},
  \citenamefont {Das}, \citenamefont {Kim}, \citenamefont {Baek}, \citenamefont
  {Eom},\ and\ \citenamefont {Ramesh}}]{ZhaoT2006}%
  \BibitemOpen
  \bibfield  {author} {\bibinfo {author} {\bibfnamefont {T.}~\bibnamefont
  {Zhao}}, \bibinfo {author} {\bibfnamefont {A.}~\bibnamefont {Scholl}},
  \bibinfo {author} {\bibfnamefont {F.}~\bibnamefont {Zavaliche}}, \bibinfo
  {author} {\bibfnamefont {K.}~\bibnamefont {Lee}}, \bibinfo {author}
  {\bibfnamefont {M.}~\bibnamefont {Barry}}, \bibinfo {author} {\bibfnamefont
  {A.}~\bibnamefont {Doran}}, \bibinfo {author} {\bibfnamefont {M.~P.}\
  \bibnamefont {Cruz}}, \bibinfo {author} {\bibfnamefont {Y.~H.}\ \bibnamefont
  {Chu}}, \bibinfo {author} {\bibfnamefont {C.}~\bibnamefont {Ederer}},
  \bibinfo {author} {\bibfnamefont {N.~A.}\ \bibnamefont {Spaldin}}, \bibinfo
  {author} {\bibfnamefont {R.~R.}\ \bibnamefont {Das}}, \bibinfo {author}
  {\bibfnamefont {D.~M.}\ \bibnamefont {Kim}}, \bibinfo {author} {\bibfnamefont
  {S.~H.}\ \bibnamefont {Baek}}, \bibinfo {author} {\bibfnamefont {C.~B.}\
  \bibnamefont {Eom}},\ and\ \bibinfo {author} {\bibfnamefont {R.}~\bibnamefont
  {Ramesh}},\ }\bibfield  {title} {\bibinfo {title} {Electrical control of
  antiferromagnetic domains in multiferroic {BiFeO$_3$} films at room
  temperature},\ }\href@noop {} {\bibfield  {journal} {\bibinfo  {journal}
  {Nat. Mater.}\ }\textbf {\bibinfo {volume} {5}},\ \bibinfo {pages} {823}
  (\bibinfo {year} {2006})}\BibitemShut {NoStop}%
\bibitem [{\citenamefont {Heron}\ \emph {et~al.}(2014)\citenamefont {Heron},
  \citenamefont {Bosse}, \citenamefont {He}, \citenamefont {Gao}, \citenamefont
  {Trassin}, \citenamefont {Ye}, \citenamefont {Clarkson}, \citenamefont
  {Wang}, \citenamefont {Liu}, \citenamefont {Salahuddin}, \citenamefont
  {Ralph}, \citenamefont {Schlom}, \citenamefont {Iniguez}, \citenamefont
  {Huey},\ and\ \citenamefont {Ramesh}}]{HeronJT2014}%
  \BibitemOpen
  \bibfield  {author} {\bibinfo {author} {\bibfnamefont {J.~T.}\ \bibnamefont
  {Heron}}, \bibinfo {author} {\bibfnamefont {J.~L.}\ \bibnamefont {Bosse}},
  \bibinfo {author} {\bibfnamefont {Q.}~\bibnamefont {He}}, \bibinfo {author}
  {\bibfnamefont {Y.}~\bibnamefont {Gao}}, \bibinfo {author} {\bibfnamefont
  {M.}~\bibnamefont {Trassin}}, \bibinfo {author} {\bibfnamefont
  {L.}~\bibnamefont {Ye}}, \bibinfo {author} {\bibfnamefont {J.~D.}\
  \bibnamefont {Clarkson}}, \bibinfo {author} {\bibfnamefont {C.}~\bibnamefont
  {Wang}}, \bibinfo {author} {\bibfnamefont {J.}~\bibnamefont {Liu}}, \bibinfo
  {author} {\bibfnamefont {S.}~\bibnamefont {Salahuddin}}, \bibinfo {author}
  {\bibfnamefont {D.~C.}\ \bibnamefont {Ralph}}, \bibinfo {author}
  {\bibfnamefont {D.~G.}\ \bibnamefont {Schlom}}, \bibinfo {author}
  {\bibfnamefont {J.}~\bibnamefont {Iniguez}}, \bibinfo {author} {\bibfnamefont
  {B.~D.}\ \bibnamefont {Huey}},\ and\ \bibinfo {author} {\bibfnamefont
  {R.}~\bibnamefont {Ramesh}},\ }\bibfield  {title} {\bibinfo {title}
  {Deterministic switching of ferromagnetism at room temperature using an
  electric field},\ }\href@noop {} {\bibfield  {journal} {\bibinfo  {journal}
  {Nature}\ }\textbf {\bibinfo {volume} {516}},\ \bibinfo {pages} {370–373}
  (\bibinfo {year} {2014})}\BibitemShut {NoStop}%
\bibitem [{\citenamefont {Li}\ and\ \citenamefont {Birol}(2020)}]{Shutong2020}%
  \BibitemOpen
  \bibfield  {author} {\bibinfo {author} {\bibfnamefont {S.}~\bibnamefont
  {Li}}\ and\ \bibinfo {author} {\bibfnamefont {T.}~\bibnamefont {Birol}},\
  }\bibfield  {title} {\bibinfo {title} {Suppressing the ferroelectric
  switching barrier in hybrid improper ferroelectrics},\ }\href@noop {}
  {\bibfield  {journal} {\bibinfo  {journal} {Npj Comput. Mater.}\ }\textbf
  {\bibinfo {volume} {6}},\ \bibinfo {pages} {168} (\bibinfo {year}
  {2020})}\BibitemShut {NoStop}%
\bibitem [{\citenamefont {Nowadnick}\ and\ \citenamefont
  {Fennie}(2016)}]{PhysRevB.94.104105}%
  \BibitemOpen
  \bibfield  {author} {\bibinfo {author} {\bibfnamefont {E.~A.}\ \bibnamefont
  {Nowadnick}}\ and\ \bibinfo {author} {\bibfnamefont {C.~J.}\ \bibnamefont
  {Fennie}},\ }\bibfield  {title} {\bibinfo {title} {Domains and ferroelectric
  switching pathways in {Ca$_3$Ti$_2$O$_7$} from first principles},\
  }\href@noop {} {\bibfield  {journal} {\bibinfo  {journal} {Phys. Rev. B}\
  }\textbf {\bibinfo {volume} {94}},\ \bibinfo {pages} {104105} (\bibinfo
  {year} {2016})}\BibitemShut {NoStop}%
\bibitem [{\citenamefont {Pokhrel}\ and\ \citenamefont
  {Nowadnick}(2023)}]{PhysRevB.107.054108}%
  \BibitemOpen
  \bibfield  {author} {\bibinfo {author} {\bibfnamefont {N.}~\bibnamefont
  {Pokhrel}}\ and\ \bibinfo {author} {\bibfnamefont {E.~A.}\ \bibnamefont
  {Nowadnick}},\ }\bibfield  {title} {\bibinfo {title} {Ferroelectric switching
  pathways and domain structure of {SrBi$_2$(Ta,Nb)$_2$O$_9$} from first
  principles},\ }\href@noop {} {\bibfield  {journal} {\bibinfo  {journal}
  {Phys. Rev. B}\ }\textbf {\bibinfo {volume} {107}},\ \bibinfo {pages}
  {054108} (\bibinfo {year} {2023})}\BibitemShut {NoStop}%
\bibitem [{\citenamefont {Zhou}\ \emph {et~al.}(2022)\citenamefont {Zhou},
  \citenamefont {Dong}, \citenamefont {Shan}, \citenamefont {Ji},\ and\
  \citenamefont {Zhang}}]{PhysRevB.105.075408}%
  \BibitemOpen
  \bibfield  {author} {\bibinfo {author} {\bibfnamefont {Y.}~\bibnamefont
  {Zhou}}, \bibinfo {author} {\bibfnamefont {S.}~\bibnamefont {Dong}}, \bibinfo
  {author} {\bibfnamefont {C.}~\bibnamefont {Shan}}, \bibinfo {author}
  {\bibfnamefont {K.}~\bibnamefont {Ji}},\ and\ \bibinfo {author}
  {\bibfnamefont {J.}~\bibnamefont {Zhang}},\ }\bibfield  {title} {\bibinfo
  {title} {Two-dimensional ferroelectricity induced by octahedral rotation
  distortion in perovskite oxides},\ }\href@noop {} {\bibfield  {journal}
  {\bibinfo  {journal} {Phys. Rev. B}\ }\textbf {\bibinfo {volume} {105}},\
  \bibinfo {pages} {075408} (\bibinfo {year} {2022})}\BibitemShut {NoStop}%
\bibitem [{\citenamefont {Inzani}\ \emph {et~al.}(2022)\citenamefont {Inzani},
  \citenamefont {Pokhrel}, \citenamefont {Leclerc}, \citenamefont {Clemens},
  \citenamefont {Ramkumar}, \citenamefont {Griffin},\ and\ \citenamefont
  {Nowadnick}}]{PhysRevB.105.054434}%
  \BibitemOpen
  \bibfield  {author} {\bibinfo {author} {\bibfnamefont {K.}~\bibnamefont
  {Inzani}}, \bibinfo {author} {\bibfnamefont {N.}~\bibnamefont {Pokhrel}},
  \bibinfo {author} {\bibfnamefont {N.}~\bibnamefont {Leclerc}}, \bibinfo
  {author} {\bibfnamefont {Z.}~\bibnamefont {Clemens}}, \bibinfo {author}
  {\bibfnamefont {S.~P.}\ \bibnamefont {Ramkumar}}, \bibinfo {author}
  {\bibfnamefont {S.~M.}\ \bibnamefont {Griffin}},\ and\ \bibinfo {author}
  {\bibfnamefont {E.~A.}\ \bibnamefont {Nowadnick}},\ }\bibfield  {title}
  {\bibinfo {title} {Manipulation of spin orientation via ferroelectric
  switching in {Fe}-doped {Bi$_2$WO$_6$} from first principles},\ }\href@noop
  {} {\bibfield  {journal} {\bibinfo  {journal} {Phys. Rev. B}\ }\textbf
  {\bibinfo {volume} {105}},\ \bibinfo {pages} {054434} (\bibinfo {year}
  {2022})}\BibitemShut {NoStop}%
\bibitem [{\citenamefont {Lu}\ \emph {et~al.}(2023)\citenamefont {Lu},
  \citenamefont {Zhang}, \citenamefont {Zhou}, \citenamefont {Zhu},
  \citenamefont {Xiang}, \citenamefont {Dong}, \citenamefont {Kageyama},\ and\
  \citenamefont {Rondinelli}}]{Lu2023}%
  \BibitemOpen
  \bibfield  {author} {\bibinfo {author} {\bibfnamefont {X.-Z.}\ \bibnamefont
  {Lu}}, \bibinfo {author} {\bibfnamefont {H.-M.}\ \bibnamefont {Zhang}},
  \bibinfo {author} {\bibfnamefont {Y.}~\bibnamefont {Zhou}}, \bibinfo {author}
  {\bibfnamefont {T.}~\bibnamefont {Zhu}}, \bibinfo {author} {\bibfnamefont
  {H.}~\bibnamefont {Xiang}}, \bibinfo {author} {\bibfnamefont
  {S.}~\bibnamefont {Dong}}, \bibinfo {author} {\bibfnamefont {H.}~\bibnamefont
  {Kageyama}},\ and\ \bibinfo {author} {\bibfnamefont {J.~M.}\ \bibnamefont
  {Rondinelli}},\ }\bibfield  {title} {\bibinfo {title} {Out-of-plane
  ferroelectricity and robust magnetoelectricity in quasi-two-dimensional
  materials},\ }\href@noop {} {\bibfield  {journal} {\bibinfo  {journal} {Sci.
  Adv.}\ }\textbf {\bibinfo {volume} {9}},\ \bibinfo {pages} {eadi0138}
  (\bibinfo {year} {2023})}\BibitemShut {NoStop}%
\bibitem [{\citenamefont {Zhu}\ \emph {et~al.}(2024{\natexlab{a}})\citenamefont
  {Zhu}, \citenamefont {Lu}, \citenamefont {Aoyama}, \citenamefont {Fujita},
  \citenamefont {Nambu}, \citenamefont {Saito}, \citenamefont {Takatsu},
  \citenamefont {Kawasaki}, \citenamefont {Terauchi}, \citenamefont {Kurosawa},
  \citenamefont {Yamaji}, \citenamefont {Li}, \citenamefont {Tassel},
  \citenamefont {Ohgushi}, \citenamefont {Rondinelli},\ and\ \citenamefont
  {Kageyama}}]{Zhu2024}%
  \BibitemOpen
  \bibfield  {author} {\bibinfo {author} {\bibfnamefont {T.}~\bibnamefont
  {Zhu}}, \bibinfo {author} {\bibfnamefont {X.-Z.}\ \bibnamefont {Lu}},
  \bibinfo {author} {\bibfnamefont {T.}~\bibnamefont {Aoyama}}, \bibinfo
  {author} {\bibfnamefont {K.}~\bibnamefont {Fujita}}, \bibinfo {author}
  {\bibfnamefont {Y.}~\bibnamefont {Nambu}}, \bibinfo {author} {\bibfnamefont
  {T.}~\bibnamefont {Saito}}, \bibinfo {author} {\bibfnamefont
  {H.}~\bibnamefont {Takatsu}}, \bibinfo {author} {\bibfnamefont
  {T.}~\bibnamefont {Kawasaki}}, \bibinfo {author} {\bibfnamefont
  {T.}~\bibnamefont {Terauchi}}, \bibinfo {author} {\bibfnamefont
  {S.}~\bibnamefont {Kurosawa}}, \bibinfo {author} {\bibfnamefont
  {A.}~\bibnamefont {Yamaji}}, \bibinfo {author} {\bibfnamefont {H.-B.}\
  \bibnamefont {Li}}, \bibinfo {author} {\bibfnamefont {C.}~\bibnamefont
  {Tassel}}, \bibinfo {author} {\bibfnamefont {K.}~\bibnamefont {Ohgushi}},
  \bibinfo {author} {\bibfnamefont {J.~M.}\ \bibnamefont {Rondinelli}},\ and\
  \bibinfo {author} {\bibfnamefont {H.}~\bibnamefont {Kageyama}},\ }\bibfield
  {title} {\bibinfo {title} {Thermal multiferroics in all-inorganic
  quasi-two-dimensional halide perovskites},\ }\href@noop {} {\bibfield
  {journal} {\bibinfo  {journal} {Nat. Mater.}\ }\textbf {\bibinfo {volume}
  {23}},\ \bibinfo {pages} {182–188} (\bibinfo {year}
  {2024}{\natexlab{a}})}\BibitemShut {NoStop}%
\bibitem [{\citenamefont {Zhou}\ \emph {et~al.}(2021)\citenamefont {Zhou},
  \citenamefont {Chen}, \citenamefont {Wu}, \citenamefont {Shen}, \citenamefont
  {Wang}, \citenamefont {Zhang},\ and\ \citenamefont {Sun}}]{ZY2021}%
  \BibitemOpen
  \bibfield  {author} {\bibinfo {author} {\bibfnamefont {Y.}~\bibnamefont
  {Zhou}}, \bibinfo {author} {\bibfnamefont {Z.}~\bibnamefont {Chen}}, \bibinfo
  {author} {\bibfnamefont {Z.}~\bibnamefont {Wu}}, \bibinfo {author}
  {\bibfnamefont {X.}~\bibnamefont {Shen}}, \bibinfo {author} {\bibfnamefont
  {J.}~\bibnamefont {Wang}}, \bibinfo {author} {\bibfnamefont {J.}~\bibnamefont
  {Zhang}},\ and\ \bibinfo {author} {\bibfnamefont {H.}~\bibnamefont {Sun}},\
  }\bibfield  {title} {\bibinfo {title} {Hybrid improper ferroelectricity and
  magnetoelectric coupling in a two-dimensional perovskite oxide},\ }\href@noop
  {} {\bibfield  {journal} {\bibinfo  {journal} {Phys. Rev. B}\ }\textbf
  {\bibinfo {volume} {103}},\ \bibinfo {pages} {224409} (\bibinfo {year}
  {2021})}\BibitemShut {NoStop}%
\bibitem [{SM()}]{SM}%
  \BibitemOpen
  \href@noop {} {\ }\bibinfo {note} {See Supplymenntal Material for relative
  energies of phases, mode-resolved analysis, phonon spectra, magnetic
  anisotropy energy, derivation of \textit{k $\cdot$ p} model and electronic
  band structures for 2D-RP structures of Ca$_3$Mn$_2$O$_7$, and structural
  energetics, strain effects, phonon spectra, and magnetic anisotropy energy
  for the strained 2D-RP structures of KAg$_2$Mn$_2$Cl$_7$, which includes Ref.
  [\cite{Bardeen1938}]}\BibitemShut {NoStop}%
\bibitem [{\citenamefont {Hatt}\ and\ \citenamefont
  {Spaldin}(2009)}]{Hatt2009}%
  \BibitemOpen
  \bibfield  {author} {\bibinfo {author} {\bibfnamefont {A.~J.}\ \bibnamefont
  {Hatt}}\ and\ \bibinfo {author} {\bibfnamefont {N.~A.}\ \bibnamefont
  {Spaldin}},\ }\bibfield  {title} {\bibinfo {title} {Strain effects on the
  electric polarization of {BiMnO$_3$}},\ }\href@noop {} {\bibfield  {journal}
  {\bibinfo  {journal} {Eur. Phys. J. B}\ }\textbf {\bibinfo {volume} {71}},\
  \bibinfo {pages} {435} (\bibinfo {year} {2009})}\BibitemShut {NoStop}%
\bibitem [{\citenamefont {Sai}\ \emph {et~al.}(2009)\citenamefont {Sai},
  \citenamefont {Fennie},\ and\ \citenamefont
  {Demkov}}]{PhysRevLett.102.107601}%
  \BibitemOpen
  \bibfield  {author} {\bibinfo {author} {\bibfnamefont {N.}~\bibnamefont
  {Sai}}, \bibinfo {author} {\bibfnamefont {C.~J.}\ \bibnamefont {Fennie}},\
  and\ \bibinfo {author} {\bibfnamefont {A.~A.}\ \bibnamefont {Demkov}},\
  }\bibfield  {title} {\bibinfo {title} {Absence of critical thickness in an
  ultrathin improper ferroelectric film},\ }\href@noop {} {\bibfield  {journal}
  {\bibinfo  {journal} {Phys. Rev. Lett.}\ }\textbf {\bibinfo {volume} {102}},\
  \bibinfo {pages} {107601} (\bibinfo {year} {2009})}\BibitemShut {NoStop}%
\bibitem [{\citenamefont {Garrity}\ \emph {et~al.}(2014)\citenamefont
  {Garrity}, \citenamefont {Rabe},\ and\ \citenamefont
  {Vanderbilt}}]{PhysRevLett.112.127601}%
  \BibitemOpen
  \bibfield  {author} {\bibinfo {author} {\bibfnamefont {K.~F.}\ \bibnamefont
  {Garrity}}, \bibinfo {author} {\bibfnamefont {K.~M.}\ \bibnamefont {Rabe}},\
  and\ \bibinfo {author} {\bibfnamefont {D.}~\bibnamefont {Vanderbilt}},\
  }\bibfield  {title} {\bibinfo {title} {Hyperferroelectrics: Proper
  ferroelectrics with persistent polarization},\ }\href@noop {} {\bibfield
  {journal} {\bibinfo  {journal} {Phys. Rev. Lett.}\ }\textbf {\bibinfo
  {volume} {112}},\ \bibinfo {pages} {127601} (\bibinfo {year}
  {2014})}\BibitemShut {NoStop}%
\bibitem [{\citenamefont {Akamatsu}\ \emph {et~al.}(2014)\citenamefont
  {Akamatsu}, \citenamefont {Fujita}, \citenamefont {Kuge}, \citenamefont
  {Sen~Gupta}, \citenamefont {Togo}, \citenamefont {Lei}, \citenamefont {Xue},
  \citenamefont {Stone}, \citenamefont {Rondinelli}, \citenamefont {Chen},
  \citenamefont {Tanaka}, \citenamefont {Gopalan},\ and\ \citenamefont
  {Tanaka}}]{Akamatsu2014}%
  \BibitemOpen
  \bibfield  {author} {\bibinfo {author} {\bibfnamefont {H.}~\bibnamefont
  {Akamatsu}}, \bibinfo {author} {\bibfnamefont {K.}~\bibnamefont {Fujita}},
  \bibinfo {author} {\bibfnamefont {T.}~\bibnamefont {Kuge}}, \bibinfo {author}
  {\bibfnamefont {A.}~\bibnamefont {Sen~Gupta}}, \bibinfo {author}
  {\bibfnamefont {A.}~\bibnamefont {Togo}}, \bibinfo {author} {\bibfnamefont
  {S.}~\bibnamefont {Lei}}, \bibinfo {author} {\bibfnamefont {F.}~\bibnamefont
  {Xue}}, \bibinfo {author} {\bibfnamefont {G.}~\bibnamefont {Stone}}, \bibinfo
  {author} {\bibfnamefont {J.~M.}\ \bibnamefont {Rondinelli}}, \bibinfo
  {author} {\bibfnamefont {L.-Q.}\ \bibnamefont {Chen}}, \bibinfo {author}
  {\bibfnamefont {I.}~\bibnamefont {Tanaka}}, \bibinfo {author} {\bibfnamefont
  {V.}~\bibnamefont {Gopalan}},\ and\ \bibinfo {author} {\bibfnamefont
  {K.}~\bibnamefont {Tanaka}},\ }\bibfield  {title} {\bibinfo {title}
  {Inversion symmetry breaking by oxygen octahedral rotations in the
  ruddlesden-popper {NaRTiO$_4$} family},\ }\href@noop {} {\bibfield  {journal}
  {\bibinfo  {journal} {Phys. Rev. Lett.}\ }\textbf {\bibinfo {volume} {112}},\
  \bibinfo {pages} {187602} (\bibinfo {year} {2014})}\BibitemShut {NoStop}%
\bibitem [{\citenamefont {Guan}\ \emph {et~al.}(2020)\citenamefont {Guan},
  \citenamefont {Hu}, \citenamefont {Shen}, \citenamefont {Xiang},
  \citenamefont {Zhong}, \citenamefont {Chu},\ and\ \citenamefont
  {Duan}}]{Guan2020}%
  \BibitemOpen
  \bibfield  {author} {\bibinfo {author} {\bibfnamefont {Z.}~\bibnamefont
  {Guan}}, \bibinfo {author} {\bibfnamefont {H.}~\bibnamefont {Hu}}, \bibinfo
  {author} {\bibfnamefont {X.}~\bibnamefont {Shen}}, \bibinfo {author}
  {\bibfnamefont {P.}~\bibnamefont {Xiang}}, \bibinfo {author} {\bibfnamefont
  {N.}~\bibnamefont {Zhong}}, \bibinfo {author} {\bibfnamefont
  {J.}~\bibnamefont {Chu}},\ and\ \bibinfo {author} {\bibfnamefont
  {C.}~\bibnamefont {Duan}},\ }\bibfield  {title} {\bibinfo {title} {Recent
  progress in two-dimensional ferroelectric materials},\ }\href@noop {}
  {\bibfield  {journal} {\bibinfo  {journal} {Adv. Electron. Mater.}\ }\textbf
  {\bibinfo {volume} {6}},\ \bibinfo {pages} {1900818} (\bibinfo {year}
  {2020})}\BibitemShut {NoStop}%
\bibitem [{\citenamefont {Ma}\ \emph {et~al.}(2021)\citenamefont {Ma},
  \citenamefont {Lyu}, \citenamefont {Hao}, \citenamefont {Zhao}, \citenamefont
  {Qian}, \citenamefont {Yan},\ and\ \citenamefont {Su}}]{Ma2021}%
  \BibitemOpen
  \bibfield  {author} {\bibinfo {author} {\bibfnamefont {X.-Y.}\ \bibnamefont
  {Ma}}, \bibinfo {author} {\bibfnamefont {H.-Y.}\ \bibnamefont {Lyu}},
  \bibinfo {author} {\bibfnamefont {K.-R.}\ \bibnamefont {Hao}}, \bibinfo
  {author} {\bibfnamefont {Y.-M.}\ \bibnamefont {Zhao}}, \bibinfo {author}
  {\bibfnamefont {X.}~\bibnamefont {Qian}}, \bibinfo {author} {\bibfnamefont
  {Q.-B.}\ \bibnamefont {Yan}},\ and\ \bibinfo {author} {\bibfnamefont
  {G.}~\bibnamefont {Su}},\ }\bibfield  {title} {\bibinfo {title} {Large family
  of two-dimensional ferroelectric metals discovered via machine learning},\
  }\href@noop {} {\bibfield  {journal} {\bibinfo  {journal} {Sci. Bull.}\
  }\textbf {\bibinfo {volume} {66}},\ \bibinfo {pages} {233} (\bibinfo {year}
  {2021})}\BibitemShut {NoStop}%
\bibitem [{\citenamefont {Chen}\ \emph {et~al.}(2021)\citenamefont {Chen},
  \citenamefont {Sun}, \citenamefont {Wei}, \citenamefont {Liu}, \citenamefont
  {Wen}, \citenamefont {Tian}, \citenamefont {Li},\ and\ \citenamefont
  {Chen}}]{ChenBuHang2021}%
  \BibitemOpen
  \bibfield  {author} {\bibinfo {author} {\bibfnamefont {B.~H.}\ \bibnamefont
  {Chen}}, \bibinfo {author} {\bibfnamefont {T.~L.}\ \bibnamefont {Sun}},
  \bibinfo {author} {\bibfnamefont {L.~Y.}\ \bibnamefont {Wei}}, \bibinfo
  {author} {\bibfnamefont {X.~Q.}\ \bibnamefont {Liu}}, \bibinfo {author}
  {\bibfnamefont {W.}~\bibnamefont {Wen}}, \bibinfo {author} {\bibfnamefont
  {H.}~\bibnamefont {Tian}}, \bibinfo {author} {\bibfnamefont {J.~Y.}\
  \bibnamefont {Li}},\ and\ \bibinfo {author} {\bibfnamefont {X.~M.}\
  \bibnamefont {Chen}},\ }\bibfield  {title} {\bibinfo {title} {Enhanced hybrid
  improper ferroelectricity in {Fe/Nb} cosubstituted {Ca$_3$Mn$_2$O$_7$}
  ceramics},\ }\href@noop {} {\bibfield  {journal} {\bibinfo  {journal} {J. Am.
  Ceram.}\ }\textbf {\bibinfo {volume} {104}},\ \bibinfo {pages} {4000}
  (\bibinfo {year} {2021})}\BibitemShut {NoStop}%
\bibitem [{\citenamefont {Benedek}\ and\ \citenamefont
  {Fennie}(2011{\natexlab{b}})}]{Benedek2011}%
  \BibitemOpen
  \bibfield  {author} {\bibinfo {author} {\bibfnamefont {N.~A.}\ \bibnamefont
  {Benedek}}\ and\ \bibinfo {author} {\bibfnamefont {C.~J.}\ \bibnamefont
  {Fennie}},\ }\bibfield  {title} {\bibinfo {title} {Hybrid improper
  ferroelectricity: A mechanism for controllable polarization-magnetization
  coupling},\ }\href@noop {} {\bibfield  {journal} {\bibinfo  {journal} {Phys.
  Rev. Lett.}\ }\textbf {\bibinfo {volume} {106}},\ \bibinfo {pages} {107204}
  (\bibinfo {year} {2011}{\natexlab{b}})}\BibitemShut {NoStop}%
\bibitem [{\citenamefont {Senn}\ \emph {et~al.}(2015)\citenamefont {Senn},
  \citenamefont {Bombardi}, \citenamefont {Murray}, \citenamefont {Vecchini},
  \citenamefont {Scherillo}, \citenamefont {Luo},\ and\ \citenamefont
  {Cheong}}]{PhysRevLett.114.035701}%
  \BibitemOpen
  \bibfield  {author} {\bibinfo {author} {\bibfnamefont {M.~S.}\ \bibnamefont
  {Senn}}, \bibinfo {author} {\bibfnamefont {A.}~\bibnamefont {Bombardi}},
  \bibinfo {author} {\bibfnamefont {C.~A.}\ \bibnamefont {Murray}}, \bibinfo
  {author} {\bibfnamefont {C.}~\bibnamefont {Vecchini}}, \bibinfo {author}
  {\bibfnamefont {A.}~\bibnamefont {Scherillo}}, \bibinfo {author}
  {\bibfnamefont {X.}~\bibnamefont {Luo}},\ and\ \bibinfo {author}
  {\bibfnamefont {S.~W.}\ \bibnamefont {Cheong}},\ }\bibfield  {title}
  {\bibinfo {title} {Negative thermal expansion in hybrid improper
  ferroelectric ruddlesden-popper perovskites by symmetry trapping},\
  }\href@noop {} {\bibfield  {journal} {\bibinfo  {journal} {Phys. Rev. Lett.}\
  }\textbf {\bibinfo {volume} {114}},\ \bibinfo {pages} {035701} (\bibinfo
  {year} {2015})}\BibitemShut {NoStop}%
\bibitem [{\citenamefont {Liu}\ \emph {et~al.}(2018)\citenamefont {Liu},
  \citenamefont {Zhang}, \citenamefont {Lin}, \citenamefont {Lin},
  \citenamefont {Yang}, \citenamefont {Li}, \citenamefont {Wang}, \citenamefont
  {Li}, \citenamefont {Yan}, \citenamefont {Wang}, \citenamefont {Li},
  \citenamefont {Dong},\ and\ \citenamefont {Liu}}]{LiuMeifeng2018}%
  \BibitemOpen
  \bibfield  {author} {\bibinfo {author} {\bibfnamefont {M.}~\bibnamefont
  {Liu}}, \bibinfo {author} {\bibfnamefont {Y.}~\bibnamefont {Zhang}}, \bibinfo
  {author} {\bibfnamefont {L.-F.}\ \bibnamefont {Lin}}, \bibinfo {author}
  {\bibfnamefont {L.}~\bibnamefont {Lin}}, \bibinfo {author} {\bibfnamefont
  {S.}~\bibnamefont {Yang}}, \bibinfo {author} {\bibfnamefont {X.}~\bibnamefont
  {Li}}, \bibinfo {author} {\bibfnamefont {Y.}~\bibnamefont {Wang}}, \bibinfo
  {author} {\bibfnamefont {S.}~\bibnamefont {Li}}, \bibinfo {author}
  {\bibfnamefont {Z.}~\bibnamefont {Yan}}, \bibinfo {author} {\bibfnamefont
  {X.}~\bibnamefont {Wang}}, \bibinfo {author} {\bibfnamefont {X.-G.}\
  \bibnamefont {Li}}, \bibinfo {author} {\bibfnamefont {S.}~\bibnamefont
  {Dong}},\ and\ \bibinfo {author} {\bibfnamefont {J.-M.}\ \bibnamefont
  {Liu}},\ }\bibfield  {title} {\bibinfo {title} {Direct observation of
  ferroelectricity in {Ca$_3$Mn$_2$O$_7$} and its prominent light absorption},\
  }\href@noop {} {\bibfield  {journal} {\bibinfo  {journal} {Appl. Phys.
  Lett.}\ }\textbf {\bibinfo {volume} {113}},\ \bibinfo {pages} {022902}
  (\bibinfo {year} {2018})}\BibitemShut {NoStop}%
\bibitem [{\citenamefont {\ifmmode~\check{S}\else \v{S}\fi{}mejkal}\ \emph
  {et~al.}(2022)\citenamefont {\ifmmode~\check{S}\else \v{S}\fi{}mejkal},
  \citenamefont {Sinova},\ and\ \citenamefont
  {Jungwirth}}]{PhysRevX.12.040501}%
  \BibitemOpen
  \bibfield  {author} {\bibinfo {author} {\bibfnamefont {L.}~\bibnamefont
  {\ifmmode~\check{S}\else \v{S}\fi{}mejkal}}, \bibinfo {author} {\bibfnamefont
  {J.}~\bibnamefont {Sinova}},\ and\ \bibinfo {author} {\bibfnamefont
  {T.}~\bibnamefont {Jungwirth}},\ }\bibfield  {title} {\bibinfo {title}
  {Emerging research landscape of altermagnetism},\ }\href@noop {} {\bibfield
  {journal} {\bibinfo  {journal} {Phys. Rev. X}\ }\textbf {\bibinfo {volume}
  {12}},\ \bibinfo {pages} {040501} (\bibinfo {year} {2022})}\BibitemShut
  {NoStop}%
\bibitem [{\citenamefont {Takasuna}\ \emph {et~al.}(2017)\citenamefont
  {Takasuna}, \citenamefont {Shiogai}, \citenamefont {Matsuzaka}, \citenamefont
  {Kohda}, \citenamefont {Oyama},\ and\ \citenamefont
  {Nitta}}]{PhysRevB.96.161303}%
  \BibitemOpen
  \bibfield  {author} {\bibinfo {author} {\bibfnamefont {S.}~\bibnamefont
  {Takasuna}}, \bibinfo {author} {\bibfnamefont {J.}~\bibnamefont {Shiogai}},
  \bibinfo {author} {\bibfnamefont {S.}~\bibnamefont {Matsuzaka}}, \bibinfo
  {author} {\bibfnamefont {M.}~\bibnamefont {Kohda}}, \bibinfo {author}
  {\bibfnamefont {Y.}~\bibnamefont {Oyama}},\ and\ \bibinfo {author}
  {\bibfnamefont {J.}~\bibnamefont {Nitta}},\ }\bibfield  {title} {\bibinfo
  {title} {Weak antilocalization induced by rashba spin-orbit interaction in
  layered {III-VI} compound semiconductor gase thin films},\ }\href@noop {}
  {\bibfield  {journal} {\bibinfo  {journal} {Phys. Rev. B}\ }\textbf {\bibinfo
  {volume} {96}},\ \bibinfo {pages} {161303} (\bibinfo {year}
  {2017})}\BibitemShut {NoStop}%
\bibitem [{\citenamefont {Caviglia}\ \emph {et~al.}(2010)\citenamefont
  {Caviglia}, \citenamefont {Gabay}, \citenamefont {Gariglio}, \citenamefont
  {Reyren}, \citenamefont {Cancellieri},\ and\ \citenamefont
  {Triscone}}]{PhysRevLett.104.126803}%
  \BibitemOpen
  \bibfield  {author} {\bibinfo {author} {\bibfnamefont {A.~D.}\ \bibnamefont
  {Caviglia}}, \bibinfo {author} {\bibfnamefont {M.}~\bibnamefont {Gabay}},
  \bibinfo {author} {\bibfnamefont {S.}~\bibnamefont {Gariglio}}, \bibinfo
  {author} {\bibfnamefont {N.}~\bibnamefont {Reyren}}, \bibinfo {author}
  {\bibfnamefont {C.}~\bibnamefont {Cancellieri}},\ and\ \bibinfo {author}
  {\bibfnamefont {J.-M.}\ \bibnamefont {Triscone}},\ }\bibfield  {title}
  {\bibinfo {title} {Tunable rashba spin-orbit interaction at oxide
  interfaces},\ }\href@noop {} {\bibfield  {journal} {\bibinfo  {journal}
  {Phys. Rev. Lett.}\ }\textbf {\bibinfo {volume} {104}},\ \bibinfo {pages}
  {126803} (\bibinfo {year} {2010})}\BibitemShut {NoStop}%
\bibitem [{\citenamefont {Choe}\ \emph {et~al.}(2019)\citenamefont {Choe},
  \citenamefont {Jin}, \citenamefont {Kim}, \citenamefont {Choi}, \citenamefont
  {Jo}, \citenamefont {Oh}, \citenamefont {Park}, \citenamefont {Jin},
  \citenamefont {Koo}, \citenamefont {Min}, \citenamefont {Hong}, \citenamefont
  {Lee}, \citenamefont {Baek},\ and\ \citenamefont {Yoo}}]{Choe2019}%
  \BibitemOpen
  \bibfield  {author} {\bibinfo {author} {\bibfnamefont {D.}~\bibnamefont
  {Choe}}, \bibinfo {author} {\bibfnamefont {M.-J.}\ \bibnamefont {Jin}},
  \bibinfo {author} {\bibfnamefont {S.-I.}\ \bibnamefont {Kim}}, \bibinfo
  {author} {\bibfnamefont {H.-J.}\ \bibnamefont {Choi}}, \bibinfo {author}
  {\bibfnamefont {J.}~\bibnamefont {Jo}}, \bibinfo {author} {\bibfnamefont
  {I.}~\bibnamefont {Oh}}, \bibinfo {author} {\bibfnamefont {J.}~\bibnamefont
  {Park}}, \bibinfo {author} {\bibfnamefont {H.}~\bibnamefont {Jin}}, \bibinfo
  {author} {\bibfnamefont {H.~C.}\ \bibnamefont {Koo}}, \bibinfo {author}
  {\bibfnamefont {B.-C.}\ \bibnamefont {Min}}, \bibinfo {author} {\bibfnamefont
  {S.}~\bibnamefont {Hong}}, \bibinfo {author} {\bibfnamefont {H.-W.}\
  \bibnamefont {Lee}}, \bibinfo {author} {\bibfnamefont {S.-H.}\ \bibnamefont
  {Baek}},\ and\ \bibinfo {author} {\bibfnamefont {J.-W.}\ \bibnamefont
  {Yoo}},\ }\bibfield  {title} {\bibinfo {title} {Gate-tunable giant
  nonreciprocal charge transport in noncentrosymmetric oxide interfaces},\
  }\href@noop {} {\bibfield  {journal} {\bibinfo  {journal} {Nat. Commun.}\
  }\textbf {\bibinfo {volume} {10}},\ \bibinfo {pages} {4510} (\bibinfo {year}
  {2019})}\BibitemShut {NoStop}%
\bibitem [{\citenamefont {Fedchenko}\ \emph {et~al.}(2024)\citenamefont
  {Fedchenko}, \citenamefont {Minár}, \citenamefont {Akashdeep}, \citenamefont
  {D’Souza}, \citenamefont {Vasilyev}, \citenamefont {Tkach}, \citenamefont
  {Odenbreit}, \citenamefont {Nguyen}, \citenamefont {Kutnyakhov},
  \citenamefont {Wind}, \citenamefont {Wenthaus}, \citenamefont {Scholz},
  \citenamefont {Rossnagel}, \citenamefont {Hoesch}, \citenamefont
  {Aeschlimann}, \citenamefont {Stadtmüller}, \citenamefont {Kläui},
  \citenamefont {Schönhense}, \citenamefont {Jungwirth}, \citenamefont
  {Hellenes}, \citenamefont {Jakob}, \citenamefont {Šmejkal}, \citenamefont
  {Sinova},\ and\ \citenamefont {Elmers}}]{Fedchenko2024}%
  \BibitemOpen
  \bibfield  {author} {\bibinfo {author} {\bibfnamefont {O.}~\bibnamefont
  {Fedchenko}}, \bibinfo {author} {\bibfnamefont {J.}~\bibnamefont {Minár}},
  \bibinfo {author} {\bibfnamefont {A.}~\bibnamefont {Akashdeep}}, \bibinfo
  {author} {\bibfnamefont {S.~W.}\ \bibnamefont {D’Souza}}, \bibinfo {author}
  {\bibfnamefont {D.}~\bibnamefont {Vasilyev}}, \bibinfo {author}
  {\bibfnamefont {O.}~\bibnamefont {Tkach}}, \bibinfo {author} {\bibfnamefont
  {L.}~\bibnamefont {Odenbreit}}, \bibinfo {author} {\bibfnamefont
  {Q.}~\bibnamefont {Nguyen}}, \bibinfo {author} {\bibfnamefont
  {D.}~\bibnamefont {Kutnyakhov}}, \bibinfo {author} {\bibfnamefont
  {N.}~\bibnamefont {Wind}}, \bibinfo {author} {\bibfnamefont {L.}~\bibnamefont
  {Wenthaus}}, \bibinfo {author} {\bibfnamefont {M.}~\bibnamefont {Scholz}},
  \bibinfo {author} {\bibfnamefont {K.}~\bibnamefont {Rossnagel}}, \bibinfo
  {author} {\bibfnamefont {M.}~\bibnamefont {Hoesch}}, \bibinfo {author}
  {\bibfnamefont {M.}~\bibnamefont {Aeschlimann}}, \bibinfo {author}
  {\bibfnamefont {B.}~\bibnamefont {Stadtmüller}}, \bibinfo {author}
  {\bibfnamefont {M.}~\bibnamefont {Kläui}}, \bibinfo {author} {\bibfnamefont
  {G.}~\bibnamefont {Schönhense}}, \bibinfo {author} {\bibfnamefont
  {T.}~\bibnamefont {Jungwirth}}, \bibinfo {author} {\bibfnamefont {A.~B.}\
  \bibnamefont {Hellenes}}, \bibinfo {author} {\bibfnamefont {G.}~\bibnamefont
  {Jakob}}, \bibinfo {author} {\bibfnamefont {L.}~\bibnamefont {Šmejkal}},
  \bibinfo {author} {\bibfnamefont {J.}~\bibnamefont {Sinova}},\ and\ \bibinfo
  {author} {\bibfnamefont {H.-J.}\ \bibnamefont {Elmers}},\ }\bibfield  {title}
  {\bibinfo {title} {Observation of time-reversal symmetry breaking in the band
  structure of altermagnetic {RuO$_2$}},\ }\href@noop {} {\bibfield  {journal}
  {\bibinfo  {journal} {Sci. Adv.}\ }\textbf {\bibinfo {volume} {10}},\
  \bibinfo {pages} {eadj4883} (\bibinfo {year} {2024})}\BibitemShut {NoStop}%
\bibitem [{\citenamefont {Lee}\ \emph {et~al.}(2024)\citenamefont {Lee},
  \citenamefont {Lee}, \citenamefont {Jung}, \citenamefont {Jung},
  \citenamefont {Kim}, \citenamefont {Lee}, \citenamefont {Seok}, \citenamefont
  {Kim}, \citenamefont {Park}, \citenamefont {\ifmmode~\check{S}\else
  \v{S}\fi{}mejkal}, \citenamefont {Kang},\ and\ \citenamefont
  {Kim}}]{PhysRevLett.132.036702}%
  \BibitemOpen
  \bibfield  {author} {\bibinfo {author} {\bibfnamefont {S.}~\bibnamefont
  {Lee}}, \bibinfo {author} {\bibfnamefont {S.}~\bibnamefont {Lee}}, \bibinfo
  {author} {\bibfnamefont {S.}~\bibnamefont {Jung}}, \bibinfo {author}
  {\bibfnamefont {J.}~\bibnamefont {Jung}}, \bibinfo {author} {\bibfnamefont
  {D.}~\bibnamefont {Kim}}, \bibinfo {author} {\bibfnamefont {Y.}~\bibnamefont
  {Lee}}, \bibinfo {author} {\bibfnamefont {B.}~\bibnamefont {Seok}}, \bibinfo
  {author} {\bibfnamefont {J.}~\bibnamefont {Kim}}, \bibinfo {author}
  {\bibfnamefont {B.~G.}\ \bibnamefont {Park}}, \bibinfo {author}
  {\bibfnamefont {L.}~\bibnamefont {\ifmmode~\check{S}\else \v{S}\fi{}mejkal}},
  \bibinfo {author} {\bibfnamefont {C.-J.}\ \bibnamefont {Kang}},\ and\
  \bibinfo {author} {\bibfnamefont {C.}~\bibnamefont {Kim}},\ }\bibfield
  {title} {\bibinfo {title} {Broken kramers degeneracy in altermagnetic mnte},\
  }\href@noop {} {\bibfield  {journal} {\bibinfo  {journal} {Phys. Rev. Lett.}\
  }\textbf {\bibinfo {volume} {132}},\ \bibinfo {pages} {036702} (\bibinfo
  {year} {2024})}\BibitemShut {NoStop}%
\bibitem [{\citenamefont {Wojde\l{}}\ and\ \citenamefont
  {\'I\~niguez}(2009)}]{PhysRevLett.103.267205}%
  \BibitemOpen
  \bibfield  {author} {\bibinfo {author} {\bibfnamefont {J.~C.}\ \bibnamefont
  {Wojde\l{}}}\ and\ \bibinfo {author} {\bibfnamefont {J.}~\bibnamefont
  {\'I\~niguez}},\ }\bibfield  {title} {\bibinfo {title} {Magnetoelectric
  response of multiferroic {BiFeO$_3$} and related materials from
  first-principles calculations},\ }\href@noop {} {\bibfield  {journal}
  {\bibinfo  {journal} {Phys. Rev. Lett.}\ }\textbf {\bibinfo {volume} {103}},\
  \bibinfo {pages} {267205} (\bibinfo {year} {2009})}\BibitemShut {NoStop}%
\bibitem [{\citenamefont {Madsen}\ \emph {et~al.}(2018)\citenamefont {Madsen},
  \citenamefont {Carrete},\ and\ \citenamefont {Verstraete}}]{Madsen2018}%
  \BibitemOpen
  \bibfield  {author} {\bibinfo {author} {\bibfnamefont {G.~K.~H.}\
  \bibnamefont {Madsen}}, \bibinfo {author} {\bibfnamefont {J.}~\bibnamefont
  {Carrete}},\ and\ \bibinfo {author} {\bibfnamefont {M.~J.}\ \bibnamefont
  {Verstraete}},\ }\bibfield  {title} {\bibinfo {title} {{BoltzTraP2}, a
  program for interpolating band structures and calculating semi-classical
  transport coefficients},\ }\href@noop {} {\bibfield  {journal} {\bibinfo
  {journal} {Comput. Phys. Commun.}\ }\textbf {\bibinfo {volume} {231}},\
  \bibinfo {pages} {140} (\bibinfo {year} {2018})}\BibitemShut {NoStop}%
\bibitem [{\citenamefont {Ye}\ \emph {et~al.}(2021)\citenamefont {Ye},
  \citenamefont {Wang}, \citenamefont {Bai}, \citenamefont {Zhang},
  \citenamefont {Wu}, \citenamefont {Zhang},\ and\ \citenamefont
  {Wang}}]{PhysRevB.104.075433}%
  \BibitemOpen
  \bibfield  {author} {\bibinfo {author} {\bibfnamefont {H.}~\bibnamefont
  {Ye}}, \bibinfo {author} {\bibfnamefont {X.}~\bibnamefont {Wang}}, \bibinfo
  {author} {\bibfnamefont {D.}~\bibnamefont {Bai}}, \bibinfo {author}
  {\bibfnamefont {J.}~\bibnamefont {Zhang}}, \bibinfo {author} {\bibfnamefont
  {X.}~\bibnamefont {Wu}}, \bibinfo {author} {\bibfnamefont {G.~P.}\
  \bibnamefont {Zhang}},\ and\ \bibinfo {author} {\bibfnamefont
  {J.}~\bibnamefont {Wang}},\ }\bibfield  {title} {\bibinfo {title}
  {Significant enhancement of magnetic anisotropy and conductivity in
  {GaN/CrI$_3$} van der waals heterostructures via electrostatic doping},\
  }\href@noop {} {\bibfield  {journal} {\bibinfo  {journal} {Phys. Rev. B}\
  }\textbf {\bibinfo {volume} {104}},\ \bibinfo {pages} {075433} (\bibinfo
  {year} {2021})}\BibitemShut {NoStop}%
\bibitem [{\citenamefont {Yuan}\ \emph {et~al.}(2024)\citenamefont {Yuan},
  \citenamefont {Georgescu},\ and\ \citenamefont
  {Rondinelli}}]{PhysRevLett.133.216701}%
  \BibitemOpen
  \bibfield  {author} {\bibinfo {author} {\bibfnamefont {L.-D.}\ \bibnamefont
  {Yuan}}, \bibinfo {author} {\bibfnamefont {A.~B.}\ \bibnamefont
  {Georgescu}},\ and\ \bibinfo {author} {\bibfnamefont {J.~M.}\ \bibnamefont
  {Rondinelli}},\ }\bibfield  {title} {\bibinfo {title} {Nonrelativistic spin
  splitting at the brillouin zone center in compensated magnets},\ }\href@noop
  {} {\bibfield  {journal} {\bibinfo  {journal} {Phys. Rev. Lett.}\ }\textbf
  {\bibinfo {volume} {133}},\ \bibinfo {pages} {216701} (\bibinfo {year}
  {2024})}\BibitemShut {NoStop}%
\bibitem [{\citenamefont {Liu}\ \emph {et~al.}(2025)\citenamefont {Liu},
  \citenamefont {Guo}, \citenamefont {Li},\ and\ \citenamefont
  {Liu}}]{PhysRevLett.134.116703}%
  \BibitemOpen
  \bibfield  {author} {\bibinfo {author} {\bibfnamefont {Y.}~\bibnamefont
  {Liu}}, \bibinfo {author} {\bibfnamefont {S.-D.}\ \bibnamefont {Guo}},
  \bibinfo {author} {\bibfnamefont {Y.}~\bibnamefont {Li}},\ and\ \bibinfo
  {author} {\bibfnamefont {C.-C.}\ \bibnamefont {Liu}},\ }\bibfield  {title}
  {\bibinfo {title} {Two-dimensional fully compensated ferrimagnetism},\
  }\href@noop {} {\bibfield  {journal} {\bibinfo  {journal} {Phys. Rev. Lett.}\
  }\textbf {\bibinfo {volume} {134}},\ \bibinfo {pages} {116703} (\bibinfo
  {year} {2025})}\BibitemShut {NoStop}%
\bibitem [{\citenamefont {Zhu}\ \emph {et~al.}(2024{\natexlab{b}})\citenamefont
  {Zhu}, \citenamefont {Chen}, \citenamefont {Liu}, \citenamefont {Liu},
  \citenamefont {Liu}, \citenamefont {Zha}, \citenamefont {Qu}, \citenamefont
  {Hong}, \citenamefont {Li}, \citenamefont {Jiang}, \citenamefont {Ma},
  \citenamefont {Hao}, \citenamefont {Zhu}, \citenamefont {Liu}, \citenamefont
  {Zeng}, \citenamefont {Jayaram}, \citenamefont {Lenger}, \citenamefont
  {Ding}, \citenamefont {Mo}, \citenamefont {Tanaka}, \citenamefont {Arita},
  \citenamefont {Liu}, \citenamefont {Ye}, \citenamefont {Shen}, \citenamefont
  {Wrachtrup}, \citenamefont {Huang}, \citenamefont {He}, \citenamefont {Qiao},
  \citenamefont {Liu},\ and\ \citenamefont {Liu}}]{Zhu2024-1}%
  \BibitemOpen
  \bibfield  {author} {\bibinfo {author} {\bibfnamefont {Y.-P.}\ \bibnamefont
  {Zhu}}, \bibinfo {author} {\bibfnamefont {X.}~\bibnamefont {Chen}}, \bibinfo
  {author} {\bibfnamefont {X.-R.}\ \bibnamefont {Liu}}, \bibinfo {author}
  {\bibfnamefont {Y.}~\bibnamefont {Liu}}, \bibinfo {author} {\bibfnamefont
  {P.}~\bibnamefont {Liu}}, \bibinfo {author} {\bibfnamefont {H.}~\bibnamefont
  {Zha}}, \bibinfo {author} {\bibfnamefont {G.}~\bibnamefont {Qu}}, \bibinfo
  {author} {\bibfnamefont {C.}~\bibnamefont {Hong}}, \bibinfo {author}
  {\bibfnamefont {J.}~\bibnamefont {Li}}, \bibinfo {author} {\bibfnamefont
  {Z.}~\bibnamefont {Jiang}}, \bibinfo {author} {\bibfnamefont {X.-M.}\
  \bibnamefont {Ma}}, \bibinfo {author} {\bibfnamefont {Y.-J.}\ \bibnamefont
  {Hao}}, \bibinfo {author} {\bibfnamefont {M.-Y.}\ \bibnamefont {Zhu}},
  \bibinfo {author} {\bibfnamefont {W.}~\bibnamefont {Liu}}, \bibinfo {author}
  {\bibfnamefont {M.}~\bibnamefont {Zeng}}, \bibinfo {author} {\bibfnamefont
  {S.}~\bibnamefont {Jayaram}}, \bibinfo {author} {\bibfnamefont
  {M.}~\bibnamefont {Lenger}}, \bibinfo {author} {\bibfnamefont
  {J.}~\bibnamefont {Ding}}, \bibinfo {author} {\bibfnamefont {S.}~\bibnamefont
  {Mo}}, \bibinfo {author} {\bibfnamefont {K.}~\bibnamefont {Tanaka}}, \bibinfo
  {author} {\bibfnamefont {M.}~\bibnamefont {Arita}}, \bibinfo {author}
  {\bibfnamefont {Z.}~\bibnamefont {Liu}}, \bibinfo {author} {\bibfnamefont
  {M.}~\bibnamefont {Ye}}, \bibinfo {author} {\bibfnamefont {D.}~\bibnamefont
  {Shen}}, \bibinfo {author} {\bibfnamefont {J.}~\bibnamefont {Wrachtrup}},
  \bibinfo {author} {\bibfnamefont {Y.}~\bibnamefont {Huang}}, \bibinfo
  {author} {\bibfnamefont {R.-H.}\ \bibnamefont {He}}, \bibinfo {author}
  {\bibfnamefont {S.}~\bibnamefont {Qiao}}, \bibinfo {author} {\bibfnamefont
  {Q.}~\bibnamefont {Liu}},\ and\ \bibinfo {author} {\bibfnamefont
  {C.}~\bibnamefont {Liu}},\ }\bibfield  {title} {\bibinfo {title} {Observation
  of plaid-like spin splitting in a noncoplanar antiferromagnet},\ }\href@noop
  {} {\bibfield  {journal} {\bibinfo  {journal} {Nature}\ }\textbf {\bibinfo
  {volume} {626}},\ \bibinfo {pages} {523–528} (\bibinfo {year}
  {2024}{\natexlab{b}})}\BibitemShut {NoStop}%
\bibitem [{\citenamefont {Lu}\ and\ \citenamefont {Rondinelli}(2020)}]{Lu2020}%
  \BibitemOpen
  \bibfield  {author} {\bibinfo {author} {\bibfnamefont {X.-Z.}\ \bibnamefont
  {Lu}}\ and\ \bibinfo {author} {\bibfnamefont {J.~M.}\ \bibnamefont
  {Rondinelli}},\ }\bibfield  {title} {\bibinfo {title} {Discovery principles
  and materials for symmetry-protected persistent spin textures with long spin
  lifetimes},\ }\href@noop {} {\bibfield  {journal} {\bibinfo  {journal}
  {Matter}\ }\textbf {\bibinfo {volume} {3}},\ \bibinfo {pages} {1211}
  (\bibinfo {year} {2020})}\BibitemShut {NoStop}%
\bibitem [{\citenamefont {Zhao}\ \emph {et~al.}(2020)\citenamefont {Zhao},
  \citenamefont {Nakamura}, \citenamefont {Arras}, \citenamefont {Paillard},
  \citenamefont {Chen}, \citenamefont {Gosteau}, \citenamefont {Li},
  \citenamefont {Yang},\ and\ \citenamefont
  {Bellaiche}}]{PhysRevLett.125.216405}%
  \BibitemOpen
  \bibfield  {author} {\bibinfo {author} {\bibfnamefont {H.~J.}\ \bibnamefont
  {Zhao}}, \bibinfo {author} {\bibfnamefont {H.}~\bibnamefont {Nakamura}},
  \bibinfo {author} {\bibfnamefont {R.}~\bibnamefont {Arras}}, \bibinfo
  {author} {\bibfnamefont {C.}~\bibnamefont {Paillard}}, \bibinfo {author}
  {\bibfnamefont {P.}~\bibnamefont {Chen}}, \bibinfo {author} {\bibfnamefont
  {J.}~\bibnamefont {Gosteau}}, \bibinfo {author} {\bibfnamefont
  {X.}~\bibnamefont {Li}}, \bibinfo {author} {\bibfnamefont {Y.}~\bibnamefont
  {Yang}},\ and\ \bibinfo {author} {\bibfnamefont {L.}~\bibnamefont
  {Bellaiche}},\ }\bibfield  {title} {\bibinfo {title} {Purely cubic spin
  splittings with persistent spin textures},\ }\href@noop {} {\bibfield
  {journal} {\bibinfo  {journal} {Phys. Rev. Lett.}\ }\textbf {\bibinfo
  {volume} {125}},\ \bibinfo {pages} {216405} (\bibinfo {year}
  {2020})}\BibitemShut {NoStop}%
\bibitem [{\citenamefont {Garcia}\ \emph {et~al.}(2020)\citenamefont {Garcia},
  \citenamefont {Vila}, \citenamefont {Hsu}, \citenamefont {Waintal},
  \citenamefont {Pereira},\ and\ \citenamefont
  {Roche}}]{PhysRevLett.125.256603}%
  \BibitemOpen
  \bibfield  {author} {\bibinfo {author} {\bibfnamefont {J.~H.}\ \bibnamefont
  {Garcia}}, \bibinfo {author} {\bibfnamefont {M.}~\bibnamefont {Vila}},
  \bibinfo {author} {\bibfnamefont {C.-H.}\ \bibnamefont {Hsu}}, \bibinfo
  {author} {\bibfnamefont {X.}~\bibnamefont {Waintal}}, \bibinfo {author}
  {\bibfnamefont {V.~M.}\ \bibnamefont {Pereira}},\ and\ \bibinfo {author}
  {\bibfnamefont {S.}~\bibnamefont {Roche}},\ }\bibfield  {title} {\bibinfo
  {title} {Canted persistent spin texture and quantum spin hall effect in
  {WTe$_2$}},\ }\href@noop {} {\bibfield  {journal} {\bibinfo  {journal} {Phys.
  Rev. Lett.}\ }\textbf {\bibinfo {volume} {125}},\ \bibinfo {pages} {256603}
  (\bibinfo {year} {2020})}\BibitemShut {NoStop}%
\bibitem [{\citenamefont {Lou}\ \emph {et~al.}(2020)\citenamefont {Lou},
  \citenamefont {Gu}, \citenamefont {Ji}, \citenamefont {Feng}, \citenamefont
  {Xiang},\ and\ \citenamefont {Stroppa}}]{Lou2020}%
  \BibitemOpen
  \bibfield  {author} {\bibinfo {author} {\bibfnamefont {F.}~\bibnamefont
  {Lou}}, \bibinfo {author} {\bibfnamefont {T.}~\bibnamefont {Gu}}, \bibinfo
  {author} {\bibfnamefont {J.}~\bibnamefont {Ji}}, \bibinfo {author}
  {\bibfnamefont {J.}~\bibnamefont {Feng}}, \bibinfo {author} {\bibfnamefont
  {H.}~\bibnamefont {Xiang}},\ and\ \bibinfo {author} {\bibfnamefont
  {A.}~\bibnamefont {Stroppa}},\ }\bibfield  {title} {\bibinfo {title} {Tunable
  spin textures in polar antiferromagnetic hybrid organic-inorganic perovskites
  by electric and magnetic fields},\ }\href@noop {} {\bibfield  {journal}
  {\bibinfo  {journal} {Npj Comput. Mater.}\ }\textbf {\bibinfo {volume} {6}},\
  \bibinfo {pages} {114} (\bibinfo {year} {2020})}\BibitemShut {NoStop}%
\bibitem [{\citenamefont {Bardeen}(1938)}]{Bardeen1938}%
  \BibitemOpen
  \bibfield  {author} {\bibinfo {author} {\bibfnamefont {J.}~\bibnamefont
  {Bardeen}},\ }\bibfield  {title} {\bibinfo {title} {An improved calculation
  of the energies of metallic {Li} and {Na}},\ }\href@noop {} {\bibfield
  {journal} {\bibinfo  {journal} {J. Chem. Phys.}\ }\textbf {\bibinfo {volume}
  {6}},\ \bibinfo {pages} {367} (\bibinfo {year} {1938})}\BibitemShut {NoStop}%
\end{thebibliography}%
\end{document}